\documentclass[fleqn,usenatbib]{mnras}
\usepackage{graphicx}
\usepackage{verbatim}
\usepackage{txfonts}
\usepackage{float}
\usepackage{color}
\usepackage{amsbsy}
\usepackage{gensymb}
\usepackage{textcomp}
\usepackage{dcolumn}
\usepackage{natbib}
\usepackage{multirow}
\usepackage[T1]{fontenc}
\usepackage{xspace}
\usepackage{caption}

\usepackage{tabularx}
\usepackage{comment}

\usepackage{amssymb}	
\usepackage{epsfig}
\usepackage{upgreek}
\usepackage{mathptmx}
\usepackage[utf8]{inputenc}
\usepackage{enumerate}
\usepackage{longtable,lscape}

\usepackage{grffile}
\usepackage{array,float}
\usepackage{lscape}
\usepackage{pdflscape}
\usepackage{afterpage}
\usepackage{siunitx}
\usepackage{xargs}

\usepackage[autostyle]{csquotes}
\usepackage{booktabs}
\usepackage{soul}
\usepackage{ulem}
\usepackage{siunitx}
\usepackage{lscape}

\newcolumntype{d}[1]{D{.}{\cdot}{#1}}
\definecolor{mygray}{gray}{0.6}
\newcolumntype{.}{D{.}{.}{-1}}

\bibpunct[; ]{(}{)}{,}{a}{}{;}

\newcommand{\lsun}{L$_\odot$}

\newcommand{\msun}{M$_\odot$}

\newcommand{\vlsr}{V$_{\rm{LSR}}$}

\newcommand{\mum}{$\mu$m}

\newcommand{\kms}{km\,s$^{-1}$\xspace}

\newcommand{\hii}{H\textsc{ii}}
\newcommand{\co}[2]{CO\,(#1--#2)}

\newcommand{\ks}{Kolmogorov–Smirnov}

\newcommand{\SC}{solar circle}
\newcommand{\higal}{Hi-GAL}
\newcommand{\oghres}{OGHReS}

\title[OGHReS-HiGAL Outer Galaxy Clumps]{OGHReS: Star formation in the Outer Galaxy ($\ell = 250\degr$-$280\degr$)\thanks{The full version of Tables\,\ref{tbl:guassian_parameters} and \ref{tbl:derived_clump_para} are only available in electronic form at the CDS via anonymous ftp to cdsarc.u-strasbg.fr (130.79.125.5) or via http://cdsweb.u-strasbg.fr/cgi-bin/qcat?J/MNRAS/.}}

\author[J.\,S.\,Urquhart et al.]{J.\,S.\,Urquhart,$^{1}$\thanks{E-mail: j.s.urquhart@kent.ac.uk}
C.\,K\"onig,$^{2}$ D.\,Colombo,$^{3,2}$ A.\,Karska,$^{2,3,4}$ F.\,Wyrowski,$^{2}$   K.\,M.\,Menten,$^{2}$  \newauthor  T.\,J.\,T.\,Moore,$^{5}$ J.\,Brand,$^{6}$ D.\,Elia,$^{7}$ A.\,Giannetti,$^{6}$ S.\,Leurini,$^{8}$ M.\,Figueira,$^{9,4}$ \newauthor M.-Y.\,Lee,$^{10}$   M.\,Dumke,$^{11,2}$
\\
$^{1}$ Centre for Astrophysics and Planetary Science, University of Kent, Canterbury, CT2\,7NH, UK \\
$^{2}$ Max-Planck-Institut f\"ur Radioastronomie (MPIfR), Auf dem H\"ugel 69, 53121 Bonn, Germany\\
$^{3}$ Argelander-Institut f\"ur Astronomie, Universit\"at Bonn, Auf dem H\"ugel 71, 53121 Bonn, Germany \\
$^{4}$ Institute of Astronomy, Faculty of Physics, Astronomy and Informatics, Nicolaus
Copernicus University, Grudziądzka 5, 87-100 Toruń, Poland\\
$^{5}$ Astrophysics Research Institute, Liverpool John Moores University, Liverpool Science Park, 146 Brownlow Hill, Liverpool, L3\,5RF, UK\\
$^{6}$ INAF - Istituto di Radioastronomia, Via P. Gobetti 101, I-40129 Bologna, Italy\\
$^{7}$ INAF - Istituto di Astrofisica e Planetologia Spaziali, Via Fosso del Cavaliere 100, I-00133 Roma, Italy\\
$^{8}$ INAF - Osservatorio Astronomico di Cagliari, Via della Scienza 5, I-09047 Selargius (CA), Italy\\
$^{9}$ National Centre for Nuclear Research, Pasteura 7, 02-093 Warszawa, Poland\\
$^{10}$ Korea Astronomy and Space Science Institute, 776 Daedeok-daero,
Yuseong-gu, Daejeon 34055, Republic of Korea\\
$^{11}$ Centro de Astro-Ingenier\'ia, Pontificia Universidad Cat\'olica de Chile, Av.\,Vicu\~na Mackenna 4860, Macul, Santiago, Chile
}

\date{Accepted XXX. Received YYY; in original form ZZZ}

\pubyear{2024}

\begin{document}
\label{firstpage}
\pagerange{\pageref{firstpage}--\pageref{lastpage}}
\maketitle

\begin{abstract}

We have used data from the Outer Galaxy High-Resolution Survey (OGHReS) to refine the velocities, distances, and physical properties of a large sample of 3\,584 clumps detected in far infrared/submillimetre emission in the \higal\ survey located in the $\ell =  250\degr-280\degr$ region of the Galactic plane. Using $^{12}$CO and $^{13}$CO spectra, we have determined reliable velocities to 3\,412 clumps (95\,per\,cent of the sample). In comparison to the  velocities from the \higal\ catalogue, we find good agreement for 80\,per\,cent of the sample (within 5\,\kms). Using the higher resolution and sensitivity of OGHReS has allowed us to correct the velocity for 632 clumps and provide velocities for 687 clumps for which no velocity had been previously allocated. The velocities are used with a rotation curve to refine the distances to the clumps and to calculate the clumps' properties using a distance-dependent gas-to-dust ratio. We have determined reliable physical parameters for 3\,200  outer Galaxy dense clumps ($\sim$90\,per\,cent of the \higal\ sources in the region). We find a trend of decreasing luminosity-to-mass ratio with increasing Galactocentric distance, suggesting the star formation efficiency is lower in the outer Galaxy or that it is resulting in more lower mass stars than in the inner Galaxy. We also find a similar surface density for protostellar clumps located in the inner and outer Galaxy, revealing that the surface density requirements for star formation are the same across the Galactic disc.
\end{abstract}



\section{Introduction}

Star formation is a fundamental process in astrophysics that determines the structure of galaxies and drives their evolution (\citealt{kennicutt2005}). Star formation rates are found to be very different in different kinds of galaxies; they are lower 
in irregular and dwarf galaxies than in spiral galaxies (\citealt{kennicutt2012}). Star formation rates are also found to vary significantly within galaxies due to different environmental factors  (e.g. density, metallicity, location).  If we are to understand the processes that drive galaxy evolution, it is crucial to determine the role played by environmental effects in the formation and evolution of molecular clouds, and how these, in turn, affect star formation. To do this, we need to study the properties of molecular clouds over a large range of environments.  

We are unable to study star formation in sufficient detail in external galaxies due to limited physical resolution (e.g. $>30$\,pc; \citealt{Leroy2021}). The Milky Way, however, presents a wide range of environments that are analogous to those found in other types of galaxies. In the  extreme environment of the Galactic Centre (Galactocentric distance $R_{\rm gc} < 2$\,kpc) conditions are similar to those found in star-burst galaxies (e.g. high uv-radiation flux, density, turbulence and cosmic ray flux), while the conditions found in the inner disc (2\,kpc $ <R_{\rm gc} <  8.15$\,kpc), are similar to those in nearby spiral galaxies (\citealt{kennicutt2012}). The low-density and low metallicity environment found in the outer Galaxy ($R_{\rm gc} > 8.15$\,kpc; \citealt{lepine2011}) are similar to what is found in irregular and dwarf galaxies, such as in our nearest neighbour galaxies in the Magellanic system (\citealt{sewilo2019}), and also comparable to the interstellar environments found at high redshift  (\citealt{kruijssen2013}).

Many systematic studies of molecular clouds have focused on the inner Galaxy where most of the molecular gas is located (e.g. CO Heterodyne Inner Milky Way Plane Survey (CHIMPS: \citealt{rigby2016}), Structure, Excitation and Dynamics of the Inner Galactic Interstellar Medium (SEDIGISM: \citealt{schuller2017, schuller2021}), and the Galactic Ring Survey (GRS: \citealt{jackson2006})). Detailed analysis of these data have quantified the structure, distribution, and properties of molecular clouds located in the inner Galaxy (e.g. \citealt{roman-duval2010, rigby2019, cabral2021, colombo2022}). The outer Galaxy provides a unique opportunity to study the connection between the spiral arms, molecular gas, and star formation under conditions that are very different from those found at smaller Galactocentric radii, namely lower \ion{H}{I} density (\citealt{kalberla2008}), less intense uv-radiation fields (\citealt{mathis1983}), smaller cosmic-ray fluxes (e.g. \citealt{bloemen1984}), lower metallicity (e.g. \citealt{rudolph1997}), and higher gas-to-dust ratio (\citealt{giannetti2017}). It is also where kinematics allow the arms and molecular material to be more reliably matched, unlike the significant overlap in velocity of spiral arm segments located in the inner Galaxy. 

The Outer Galaxy High Resolution Survey (OGHReS) is a systematic high-resolution (i.e. $\theta_{\rm FWHM}\approx 30$\,arcsecs) survey of the southern outer Galactic plane in $^{12}$\co{2}{1}\ and $^{13}$\co{2}{1}, mapping 100\,deg$^2$ between $180\degree < \ell < 280\degree$ and 1\degr\ in $b$ using the Atacama Pathfinder Experiment 12-m submillimeter telescope (APEX;  \citealt{gusten2006}). A detailed description of the survey and data processing will be presented in a forthcoming paper (K\"onig et al. in prep.). This survey provides a 16-fold increase in angular resolution compared to the only other unbiased CO survey of this region (i.e. by \citealt{dame2001}). It will produce a detailed census of thousands of molecular clouds and filaments (\citealt{colombo2021}) and identify any significant spurs and/or connecting bridges present in this part of the Galaxy. 

\begin{figure}
    \centering
    \includegraphics[width = 0.49\textwidth, trim=20 0 20 0]{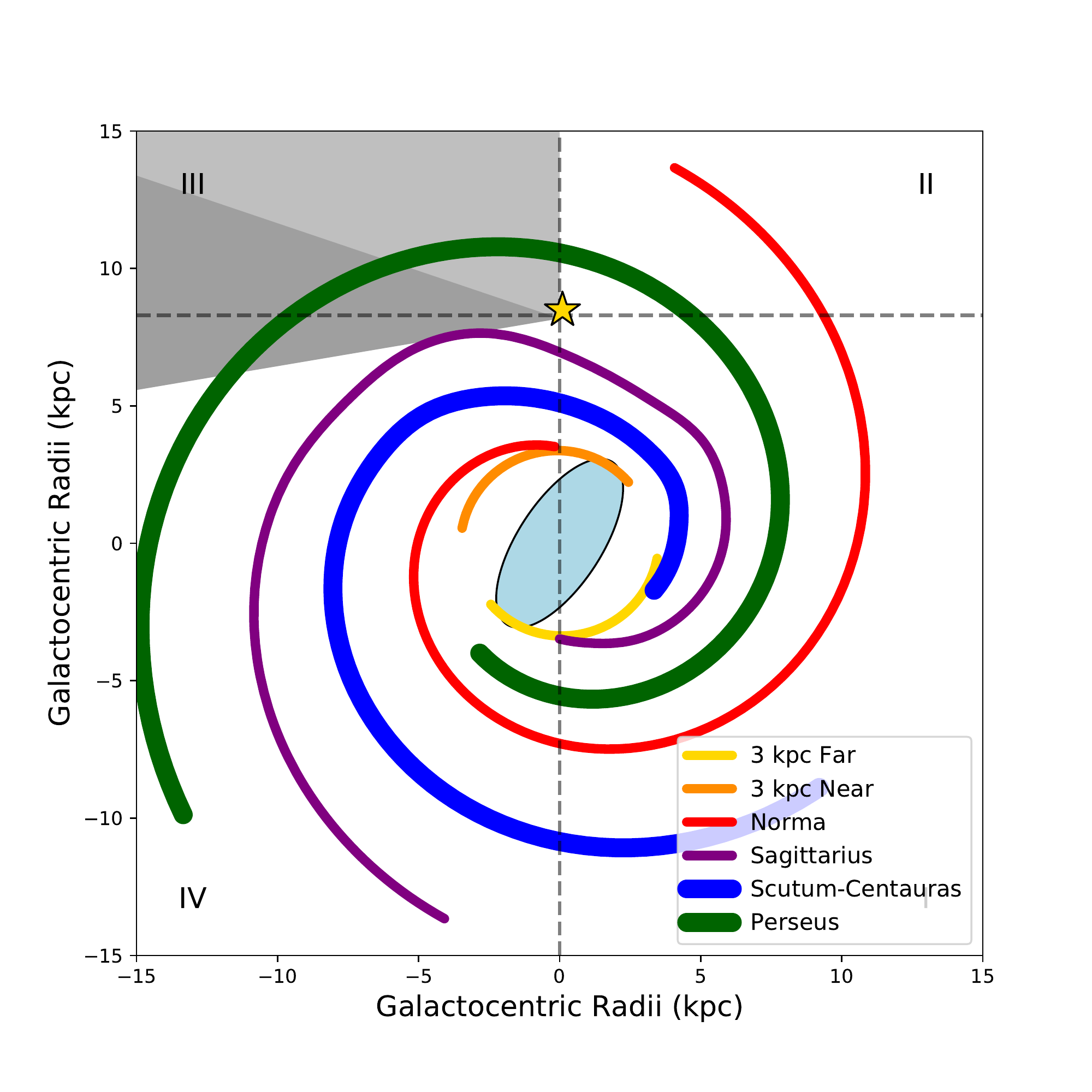}
    \caption{Schematic showing the loci of the spiral arms according to the model of \citet{taylor1993} and updated by \citet{cordes2004}, with an additional bisymmetric pair of arm segments added to represent the 3\,kpc arms. The grey shaded region indicates the coverage of the OGHReS survey while the darker grey shaded region shows the longitude coverage of the region studied in this work (see text for details). The star shows the position of the Sun and the Roman numerals identify the Galactic quadrants. The light blue oval feature located towards the centre of the diagram shows the position and orientation of the Galactic bar.}
    \label{fig:coverage_map}
\end{figure}

In this paper, we use data from a region of the survey for which the observations and data processing have been completed and the final data products are considered to be of science quality: $250\degr < \ell < 280\degr$ and $-2\degr < b < -1\degr$.  In Figure\,\ref{fig:coverage_map}, we show the full coverage of the Galactic plane by OGHReS (grey shading) and the region covered in this work (dark grey shading). We will use these data to check the velocities and distances assigned to the \higal\ clumps located in this part of the Galaxy (\citealt{mege2021}), and refine their physical properties (\citealt{elia2021}). This step is crucial for understanding how different environmental conditions affect star formation, not just in the Milky Way but also in nearby galaxies.


\section{Refining the Hi-GAL catalogue}
\label{sect:higal_cat_refining}

The region discussed in this paper has been mapped in the continuum at far-infrared and submillimetre wavelengths by \textit{Herschel} as part of the \higal\ legacy survey (\citealt{molinari2010a}). A catalogue  of $\sim$150\,000 sources covering the whole of the Galactic disc has been produced (\citealt{elia2021}). This catalogue is divided into high reliability and low reliability catalogues that consist of 94\,604 and 55\,619 clumps, respectively. The high-resolution spectroscopic data now available from OGHReS are ideally suited to complementing that survey. Using the kinematic information provided by the CO spectra, we are able to determine the velocities and distances of a large number of clumps, many for the first time, as well as to refine the velocities and distances for many of the \higal\ sources where a value had been determined.

There are 3\,584 \higal\ clumps located within the OGHReS region, 2\,073 of which are flagged as being highly reliable and 1\,511 considered to be lower reliability. The main difference between these two reliability types is that those identified with high-reliability have properties calculated through a spectral-energy distribution (SED) fit to the four \higal\ flux bands between 160\,\mum\ and 500\,\mum, while those identified with low-reliability have properties calculated through a fit to only three \higal\ flux bands in the same wavelength range, and consequently are considered to be less well constrained  (\citealt{elia2021}). 

\begin{figure*}
	\centering
        \includegraphics[width=0.45\textwidth]{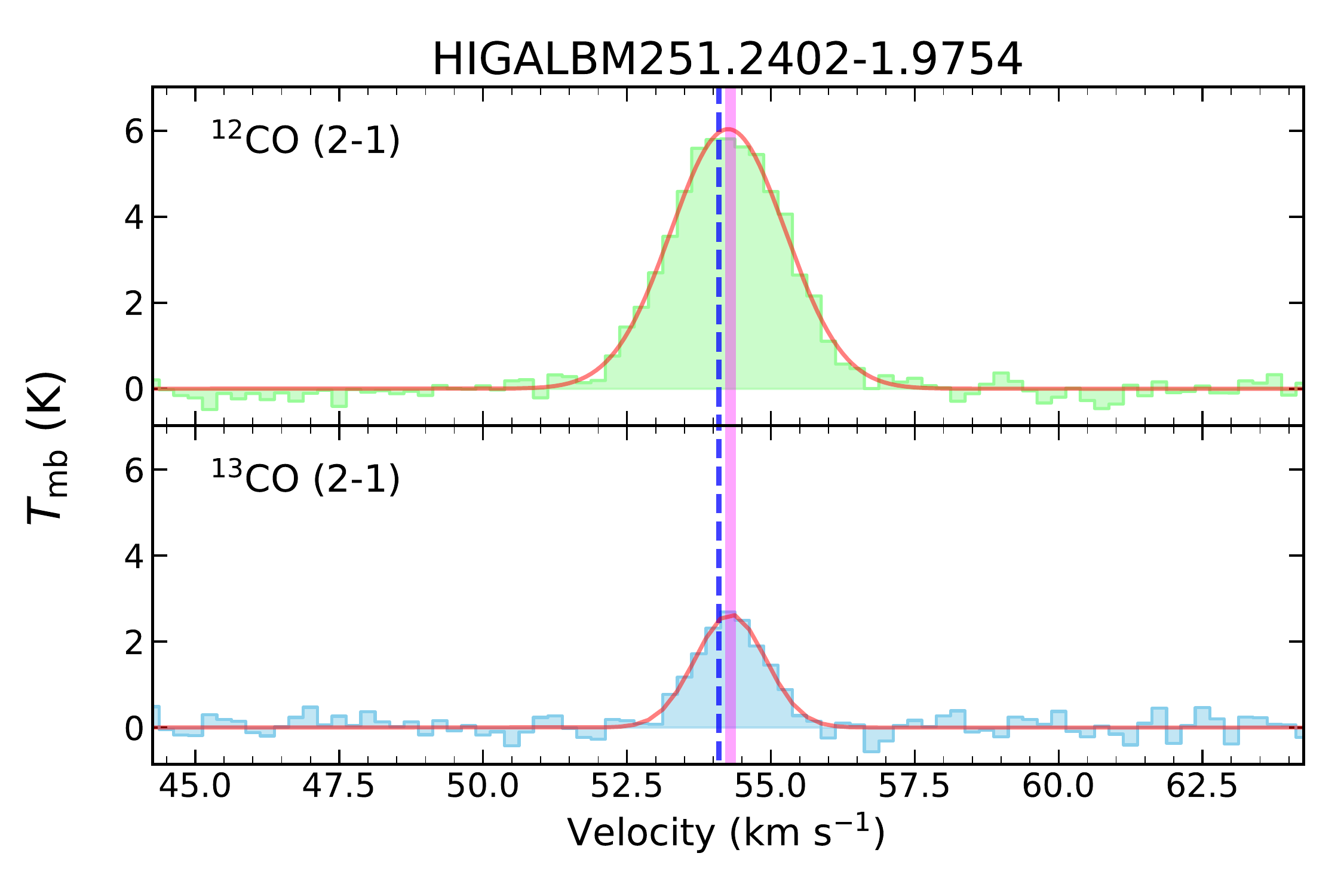}
        \includegraphics[width=0.45\textwidth]{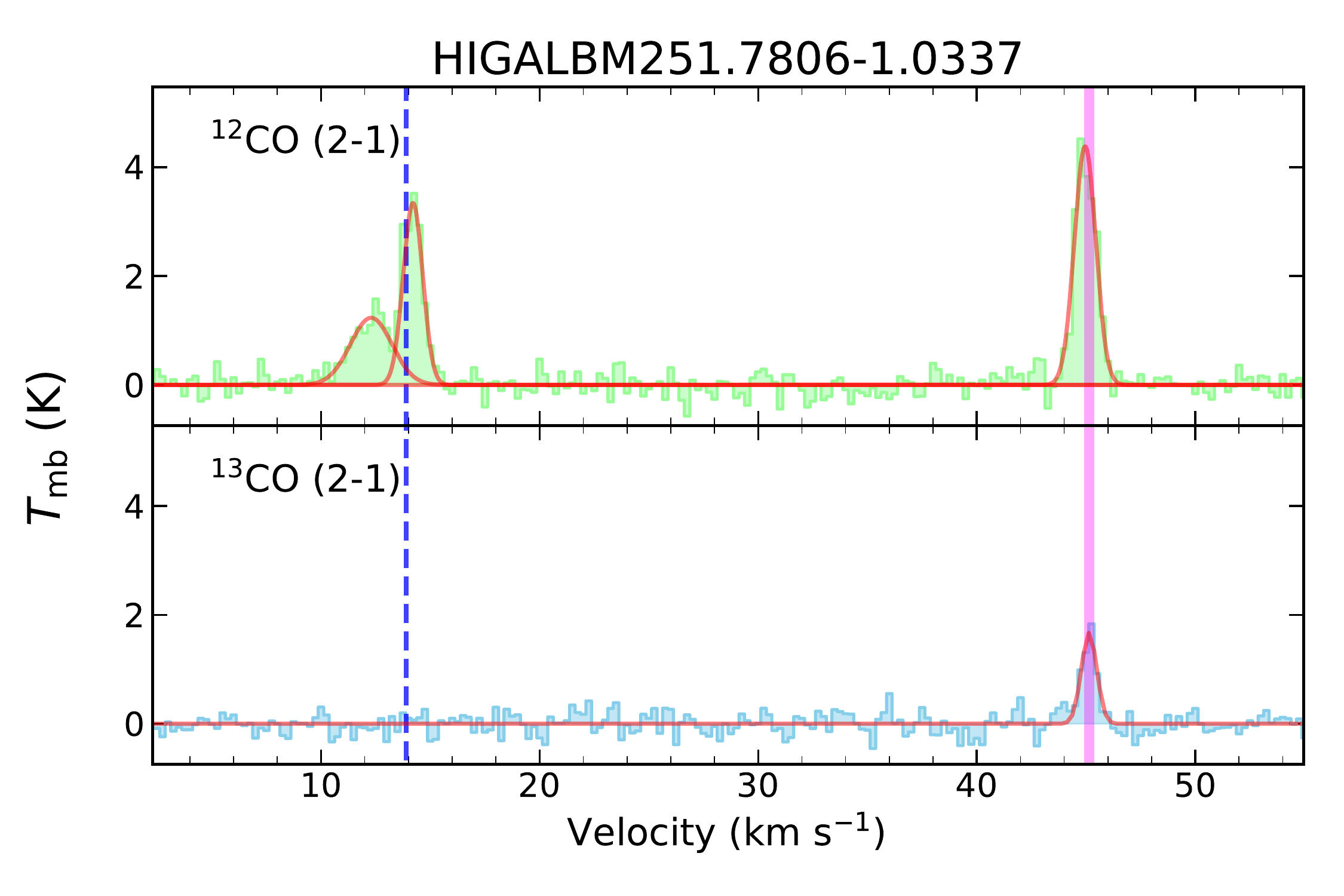}\\
        \includegraphics[width=0.45\textwidth]{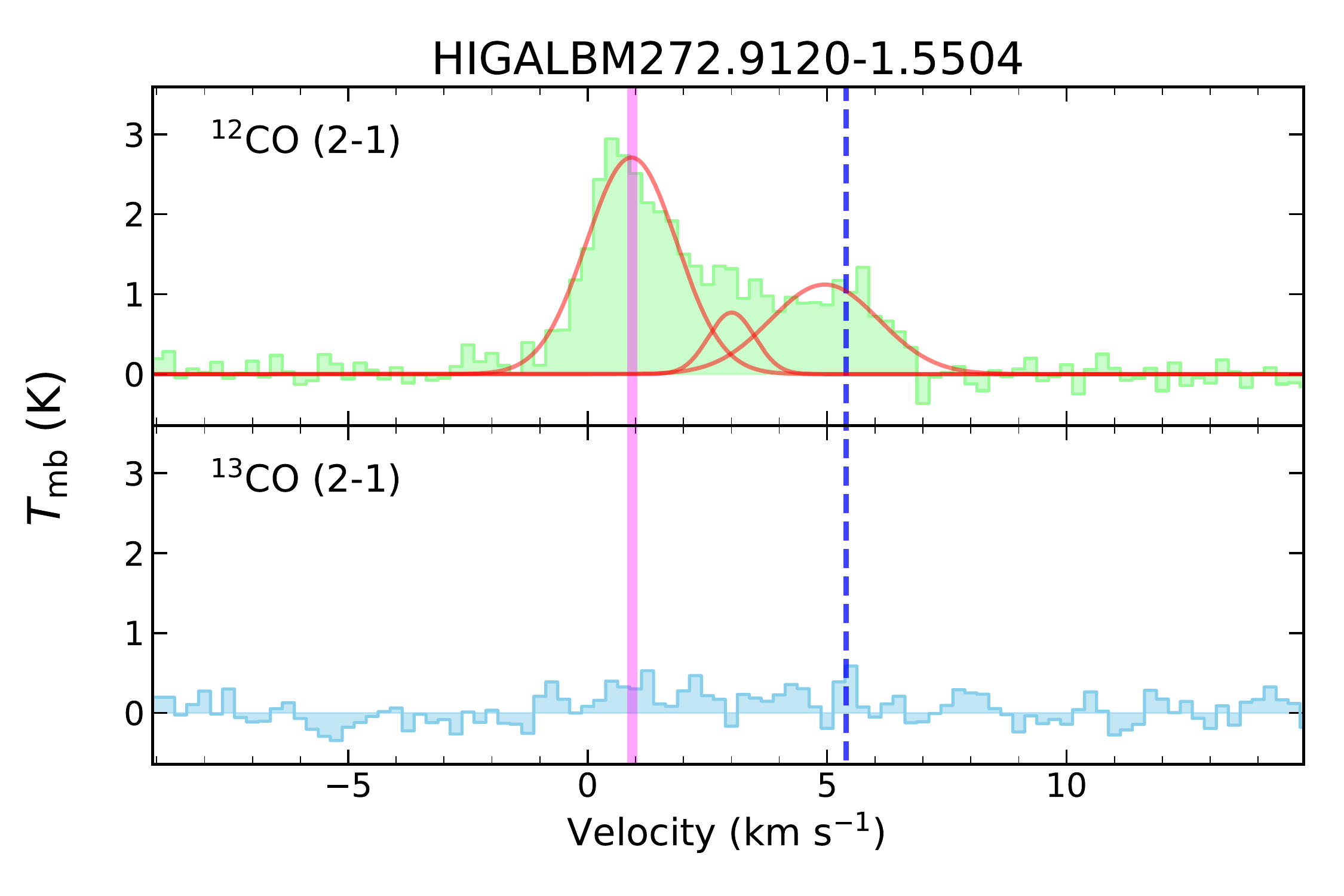}
        \includegraphics[width=0.45\textwidth]{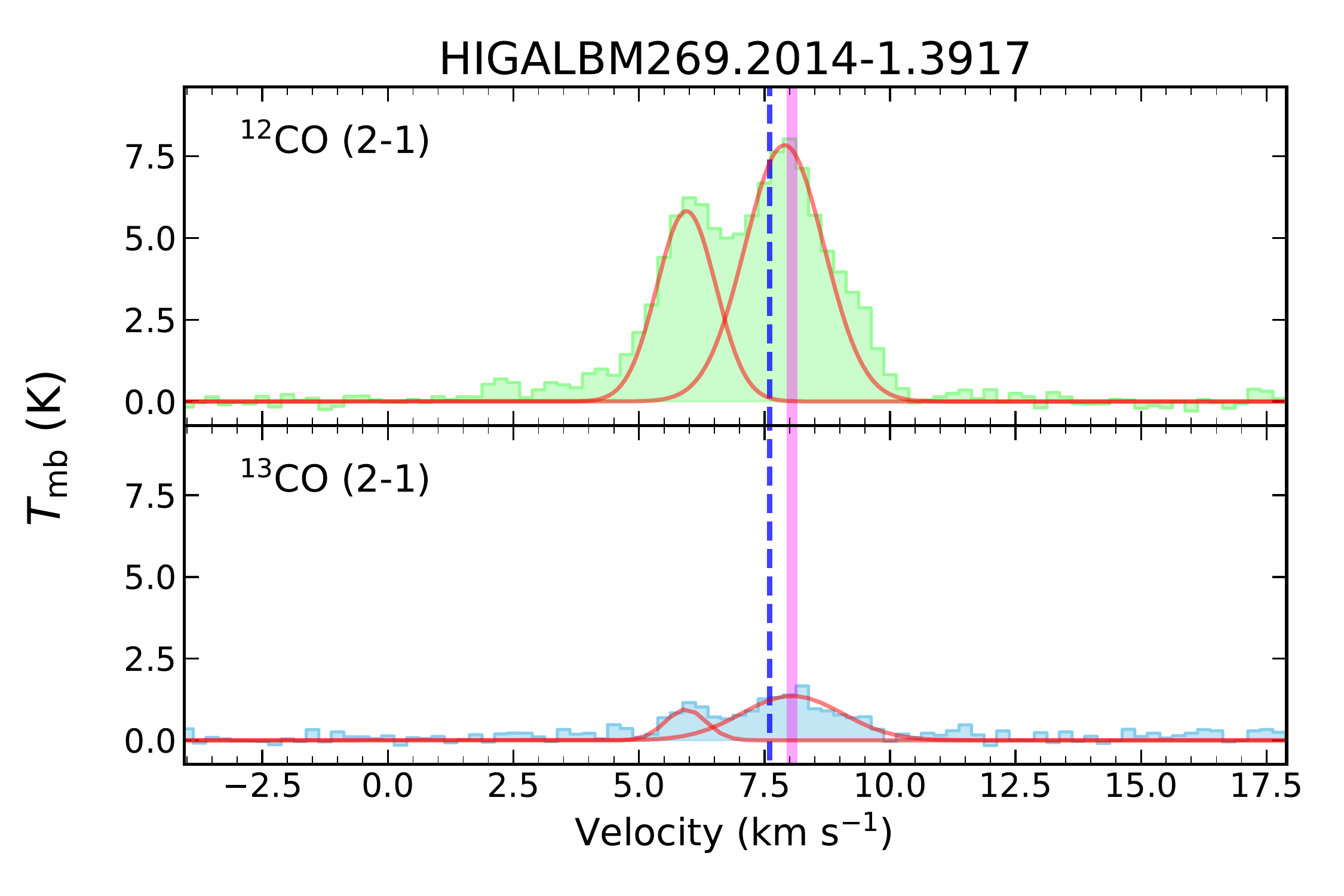}\\
        \includegraphics[width=0.45\textwidth]{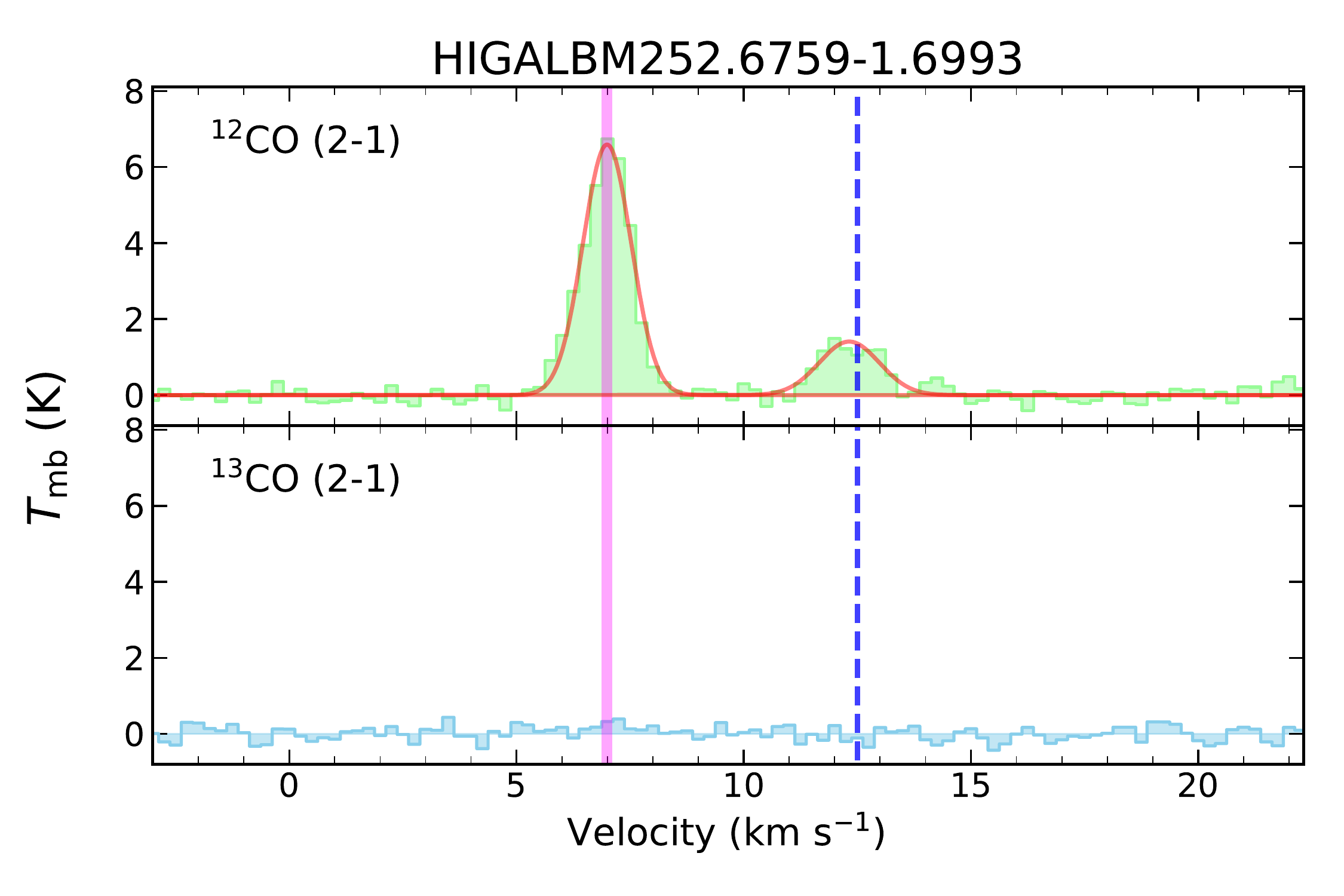}
        \includegraphics[width=0.45\textwidth]{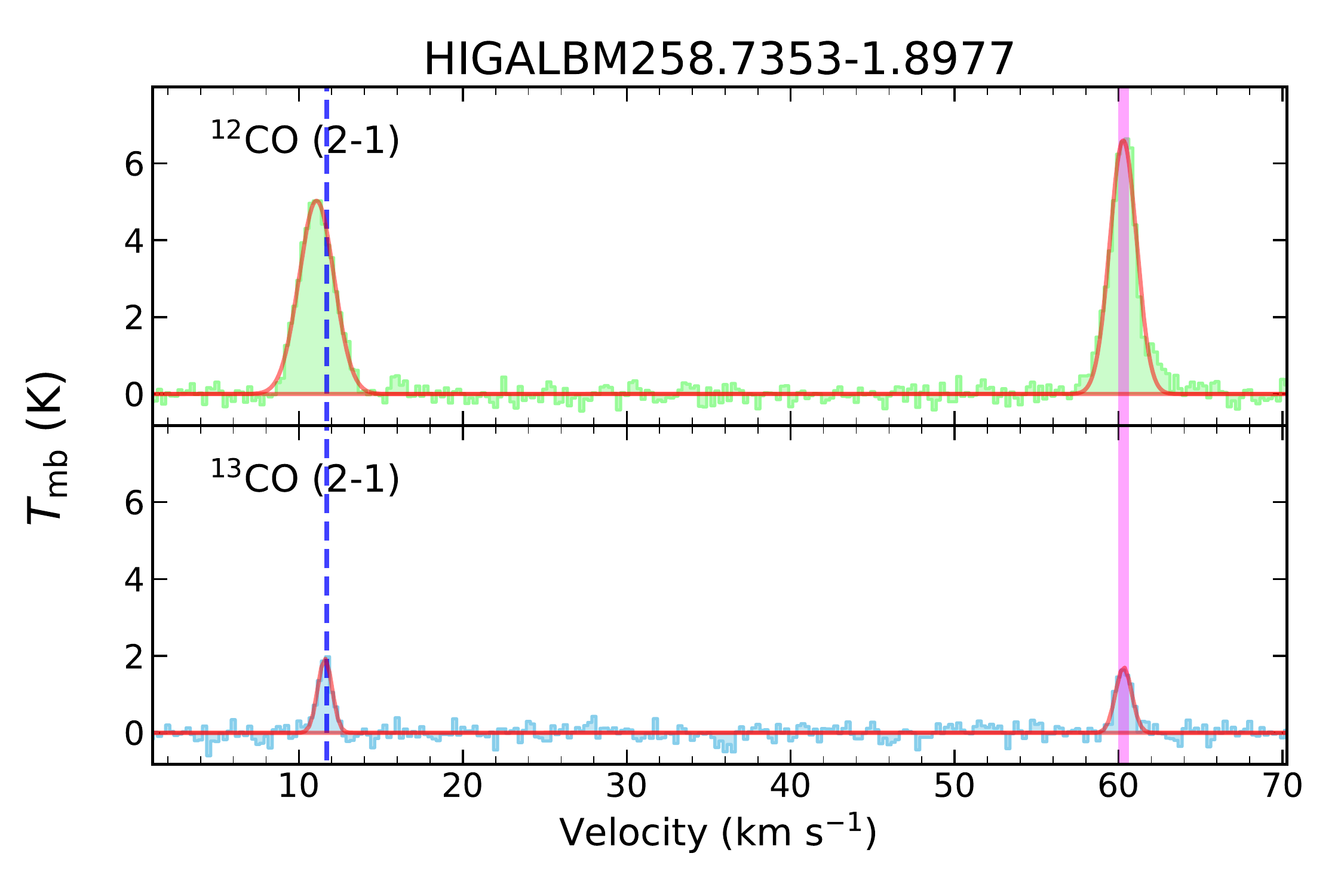}
    	\caption{CO\,(2-1) line emission toward selected \higal\ clumps. The blue vertical dashed line shows the velocity allocated by the \higal\ team (\citealt{mege2021,elia2021}) and the thick vertical magenta line shows the velocity allocated from the OGHReS data in this paper (see text for details).}
		\label{fig:velocity_examples}
	
\end{figure*}

The \higal\ catalogue distinguishes between three different evolutionary stages of clumps: unbound, bound, and protostellar (see \citealt{elia2021} for more details and caveats). The distinction between bound and unbound is determined using Larson's third law ($M(r) = 460$\,\msun\,$(r/{\rm pc})^{1.9}$) as a threshold; clumps above this threshold are considered bound while clumps below it are considered unbound. The distinction between starless (bound and unbound) clumps and protostellar clumps is based on the presence or absence of a 70-\mum\ counterpart; clumps with a 70-\mum\ counterpart are classified as protostellar.

\subsection{Velocity determination}
\label{sec:vel}

 We  extracted $^{12}$CO and $^{13}$CO\,(2-1) spectra toward all 3\,584 clumps covered by both \higal\ and OGHReS. To increase the signal-to-noise ratio, we summed up the emission over the beam (i.e. the closest 9 pixels, given the pixel size of $9.5\times 9.5$\,arcsec$^2$). The spectra cover a velocity range from $-$50\,\kms to 150\,\kms and have a velocity resolution of 0.25\,\kms.\footnote{All velocities are measured with respect to the local standard of rest (LSR).} Subsequently, a fifth-order polynomial function is fitted to the emission-free channels to correct for variations in the spectral baseline. 

  \begin{figure*}
\centering
        \includegraphics[width=0.45\textwidth]{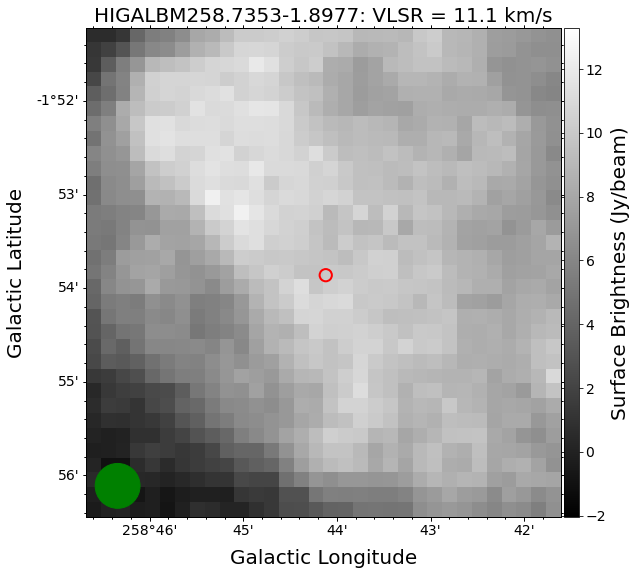}
    \includegraphics[width=0.45\textwidth]{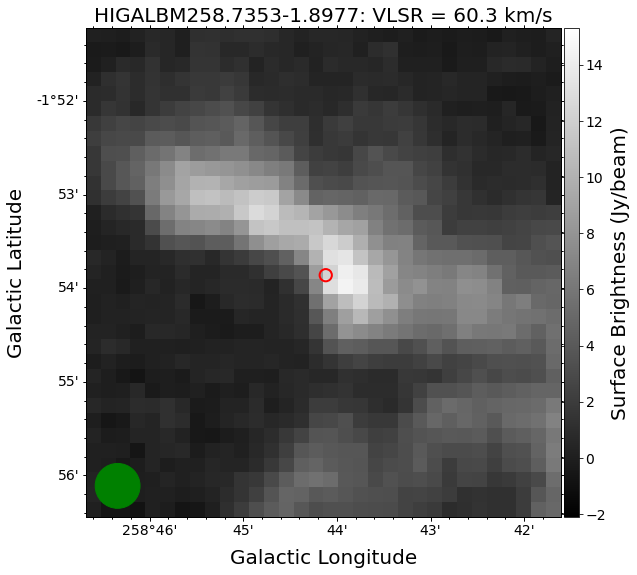}
   	\caption{Integrated emission maps for HIGALBM258.7353$-$1.8977 for the two approximately equal $^{12}$CO components detected at different velocities that make assignment of a velocity from the spectra alone unreliable (see lower right panel of Fig.\,\ref{fig:velocity_examples}). The beam size is shown as a green circle in the lower left corner and the position of the \higal\ source is indicated by the red circle. The velocity of the peak component about which the emission is integrated is given above each map. The coincidence of the \higal\ sources with compact molecular gas at $\sim$60\,\kms\ allows us to allocate this velocity to the source with a high degree of confidence.   
   	}
	\label{fig:integrated_co_maps}
	
\end{figure*}
 
 The noise is determined from the standard deviation of the emission-free channels and a 3-$\sigma$ threshold is used to search for emission in the spectra. Peaks above this threshold  are identified using the \texttt{find\_peaks} function and fitted simultaneously using the \texttt{curve\_fit} function (both functions are part of the Python \texttt{SciPy} package (\citealt{scipy}). All the fits are visually checked to ensure that their results are reliable and optimised where necessary (primarily need for blended emission components). To avoid contamination from noise spikes, we require that emission covers at least three contiguous 0.25-\kms\ channels. Furthermore, to avoid complicating the analysis, we exclude any components whose integrated intensity is less than 10\,per\,cent of the strongest component, as these are deemed to be insignificant. With this step, we make the simplifying assumption that the \higal\ clump is a distinct object at a single velocity and that their high density means that it is more likely to be associated with the most intense CO peak.

\setlength{\tabcolsep}{6pt}

\begin{table*}

\begin{center}\caption{Fitted line parameters extracted from the OGHReS $^{12}$CO and $^{13}$CO spectra.}
\label{tbl:guassian_parameters}
\begin{minipage}{\linewidth}
\small
\begin{tabular}{lc......c..c}
\hline \hline
  \multicolumn{1}{c}{}&  
  \multicolumn{4}{c}{$^{12}$CO\,(2-1)} &
  \multicolumn{1}{c}{} &
  \multicolumn{4}{c}{$^{13}$CO\,(2-1)} \\
  \cline{2-5}\cline{7-10}
  \multicolumn{1}{c}{\higal}&
  \multicolumn{1}{c}{RMS} & 
  \multicolumn{1}{c}{\vlsr} &
  \multicolumn{1}{c}{$T_{\rm mb}$}&
  \multicolumn{1}{c}{$\sigma$} &
  \multicolumn{1}{c}{} &
  \multicolumn{1}{c}{RMS} & 
  \multicolumn{1}{c}{\vlsr} &
  \multicolumn{1}{c}{$T_{\rm mb}$}&
  \multicolumn{1}{c}{$\sigma$} \\

  \multicolumn{1}{c}{name }&
  \multicolumn{1}{c}{(K)} & 
  \multicolumn{1}{c}{(\kms)} &
  \multicolumn{1}{c}{(K)}&
  \multicolumn{1}{c}{(\kms)} &
  \multicolumn{1}{c}{} &
 \multicolumn{1}{c}{(K)} & 
  \multicolumn{1}{c}{(\kms)} &
  \multicolumn{1}{c}{(K)}&
  \multicolumn{1}{c}{(\kms)}   \\
    
\hline

HIGALBM250.8892$-$1.5245	&	0.20	&	51.1	&	4.8	&	0.5	&&	0.21	&	51.2	&	1.9	&	0.3	\\
HIGALBM251.1729$-$1.9826	&	0.25	&	55.4	&	6.0	&	0.9	&&	0.25	&	55.3	&	2.1	&	0.5	\\
HIGALBM251.1801$-$1.9766	&	0.21	&	55.5	&	9.2	&	0.7	&&	0.22	&	55.5	&	2.0	&	0.5	\\
HIGALBM251.1919$-$1.9733	&	0.22	&	55.6	&	20.6	&	0.8	&&	0.23	&	55.5	&	6.7	&	0.5	\\
HIGALBM251.1931$-$1.9758	&	0.22	&	55.6	&	23.7	&	0.8	&&	0.23	&	55.5	&	7.4	&	0.6	\\
HIGALBM251.7806$-$1.0337	&	0.19	&	45.0	&	4.4	&	0.5	&&	0.19	&	45.1	&	1.7	&	0.4	\\
HIGALBM251.8034$-$1.0768	&	0.22	&	44.9	&	5.4	&	0.5	&&	0.21	&	44.9	&	2.1	&	0.2	\\
HIGALBM251.8671$-$1.8202	&	0.21	&	49.6	&	1.8	&	0.3	&&	0.18	&		&		&		\\
HIGALBM251.8908$-$1.0908	&	0.24	&	44.9	&	4.1	&	1.0	&&	0.26	&	45.2	&	1.8	&	0.5	\\
HIGALBM251.9016$-$1.0808	&	0.23	&	44.9	&	4.5	&	0.9	&&	0.23	&	45.0	&	1.6	&	0.6	\\

\hline\\
\end{tabular}\\
Notes: Only a small portion of the data is provided here. The full table is available in electronic form at the CDS via anonymous ftp to cdsarc.u-strasbg.fr (130.79.125.5) or via http://cdsweb.u-strasbg.fr/cgi-bin/qcat?J/MNRAS/.
\end{minipage}

\end{center}
\end{table*}
\setlength{\tabcolsep}{6pt}

In total, we have identified 7\,102 $^{12}$CO spectral components towards 3\,481 \higal\ clumps and 3\,009 $^{13}$CO spectral components towards 2\,584 \higal\ clumps. There are 104 clumps towards which no emission has been detected. In the following analysis we prioritise the use of the $^{13}$CO emission in allocating velocities to clumps, as it is more likely be associated with high-column-density sources, and is less affected by blending and self-absorption than the more abundant $^{12}$CO emission, which is nearly always optically thick. A single $^{13}$CO component is detected towards 2\,181 clumps, allowing a velocity to be attributed unambiguously (see the upper panels of Fig.\,\ref{fig:velocity_examples}). There are 846 clumps that are not detected in $^{13}$CO but are detected in $^{12}$CO.  Of these, there are 411 that are associated with a single $^{12}$CO emission peak; also allowing a velocity to be attributed unambiguously. Two or more $^{13}$CO and $^{12}$CO components are detected towards the remaining 785 clumps. Assigning velocities to these clumps requires additional steps, which are outlined below:



\begin{enumerate}

     \item If all significant components in a given spectrum  are close together in velocity, such that the standard deviation of the velocities is less than 5\,\kms, then the component with the highest intensity is assigned to the clump (see the middle panels of Fig.\,\ref{fig:velocity_examples}). The uncertainty due to streaming motions is of order $\pm$7\,\kms\ (\citealt{reid2014}) and this will dominate the uncertainty in the distance measurement. This criterion allows a velocity to be assigned to 257 clumps using the $^{13}$CO data and 152 clumps using the $^{12}$CO data. \\

    \item If the strongest component in a spectrum has an integrated intensity that is a factor of more than 2 times higher than that of the next strongest component in the same spectrum, then the velocity of the strongest component is assigned to the clump (see lower left panel of Fig.\,\ref{fig:velocity_examples} for an example). This criterion allows a velocity to be assigned to a further 30 clumps using the $^{13}$CO data and 107 clumps using the $^{12}$CO data.\\

    
    \item The remaining 343 clumps have two or more components of roughly equal intensity (within a factor of 2) with significantly different velocities. To resolve the velocity ambiguity in these cases we created $5\times 5$-arcmin integrated maps of the strongest components and selected the velocity component for which the peak of the emission most closely correlates with the position of the \higal\ clump (cf. \citealt{urquhart2021}). To illustrate this method, we show in Figure\,\ref{fig:integrated_co_maps} the $^{12}$CO integrated emission maps towards the \higal\ clump  HIGALBM258.7353$-$1.8977. The spectrum for this clump  shows two approximately equal intensity peaks at 11.1\,\kms\ and 60.3\,\kms\ (see lower right panel of Fig.\,\ref{fig:velocity_examples}). Emission in the integrated maps clearly favours the component at 60.3\,\kms\ that is coincident and position and morphologically correlated with the \higal\ clump. In this case, the component detected at 11.1\,\kms\ is likely to be associated with local diffuse gas. This method has been successful in allocating velocities for a further 274 clumps. 


\end{enumerate}

\begin{figure}
	\centering
        \includegraphics[width=0.45\textwidth]{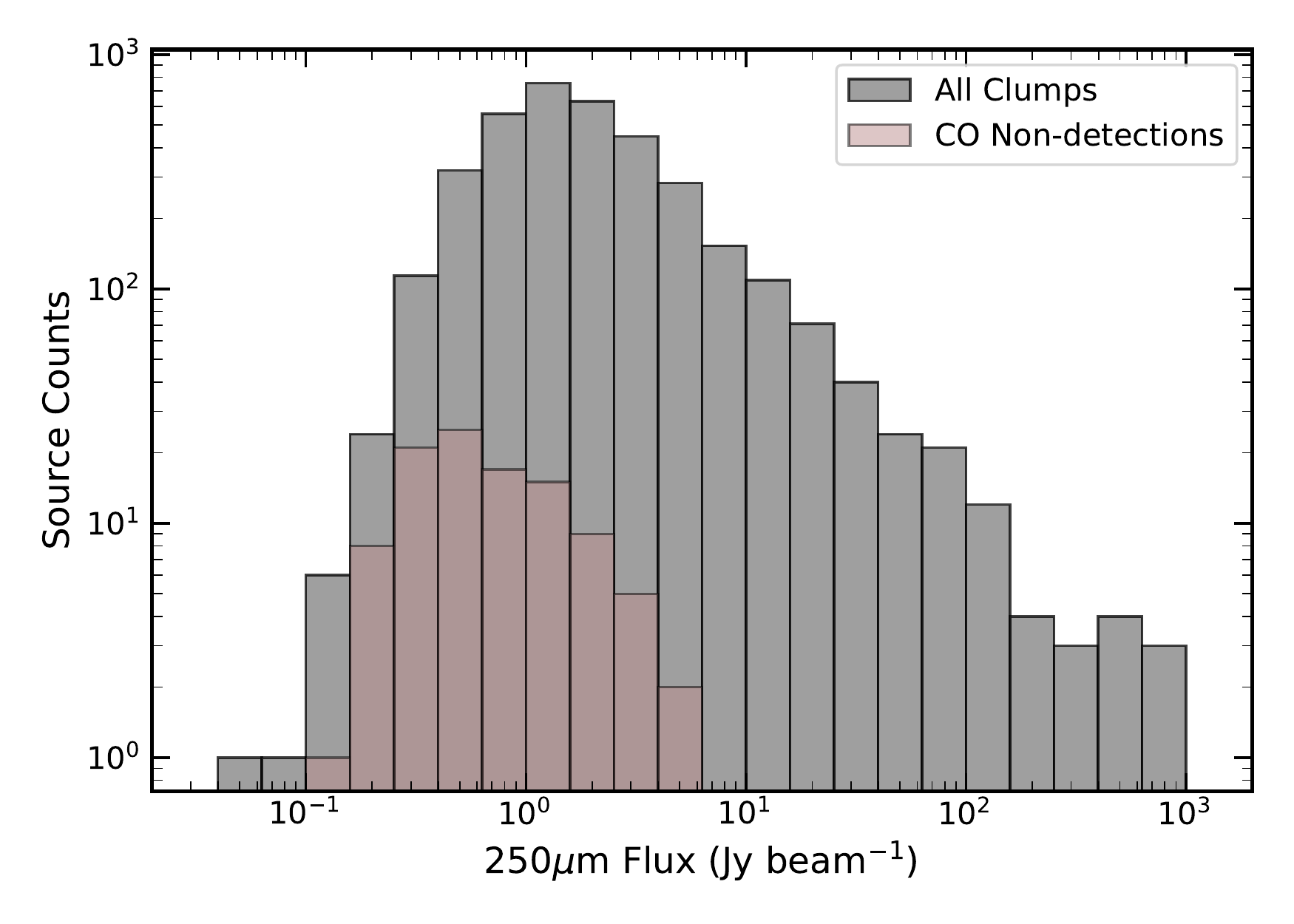}
    
    	\caption{\higal\, 250-$\mu$m flux distribution for all clumps in the OGHReS region studied here. The bin size used is 0.2\,dex.}
		\label{fig:flux_comparison}
	
\end{figure}

In total, we have determined reliable velocities for 3\,412 \higal\ clumps in the studied region (corresponding to 95\,per\,cent of the sample). In Table\,\ref{tbl:guassian_parameters}, we provide the fitted velocities, intensities and line widths of the $^{12}$CO and $^{13}$CO emission associated with the \higal\ clumps. We have been unable to allocate a reliable velocity to 69 clumps for which multiple CO components have been detected and report 104 non-detections. In Figure\,\ref{fig:flux_comparison}, we show the 250-$\mu$m flux distribution of the complete sample and the CO non-detections. This shows that the non-detections are some of the weakest in the sample.

\subsection{Comparison with velocities assigned to \higal\ clumps}

\begin{figure}
	\centering

        \includegraphics[width=0.45\textwidth]{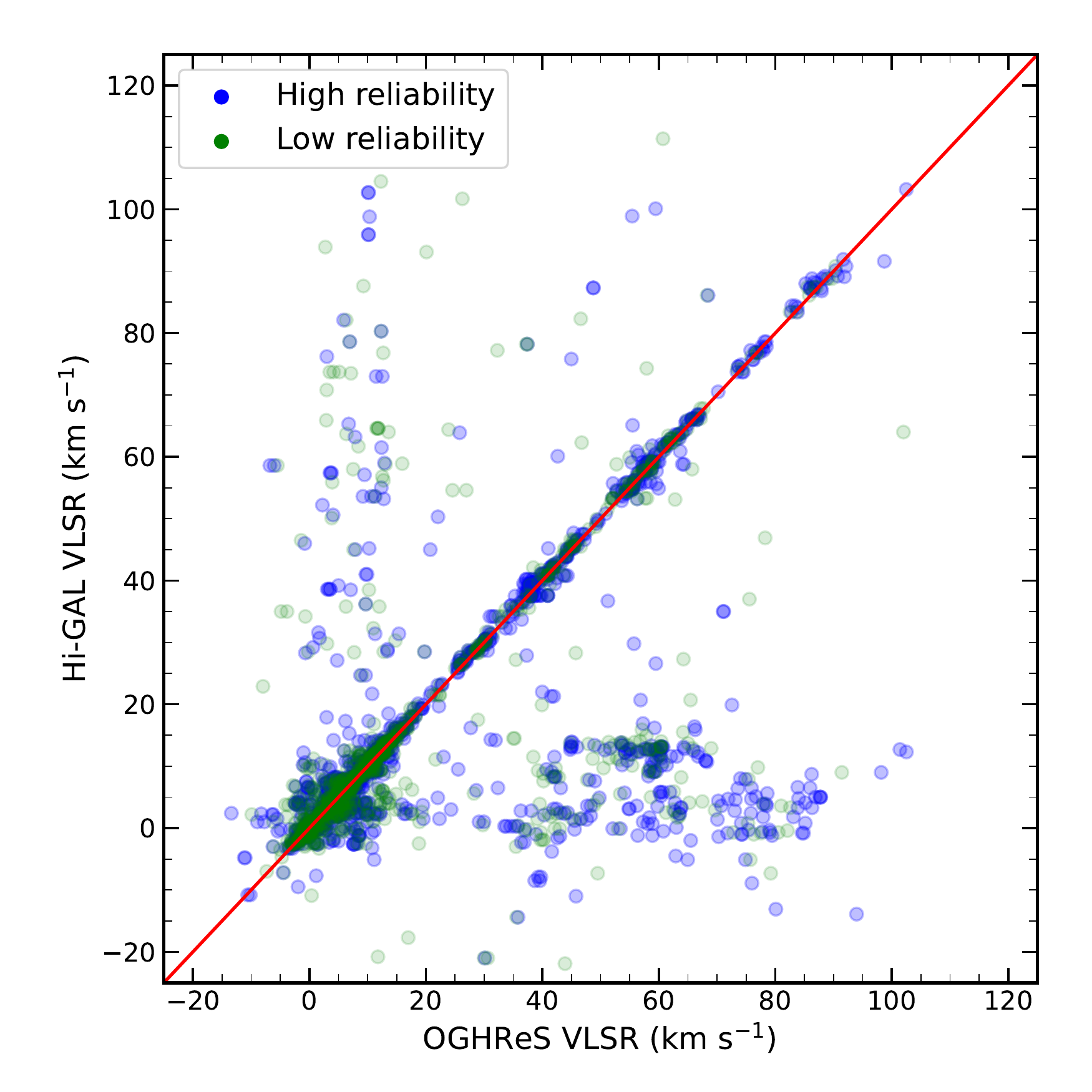}
    
    	\caption{Comparison of the velocities reported by \higal\ and the velocities determined in this paper from fits to the OGHReS spectra. The blue and green circles show the high and low reliability sources respectively, and the red line shows the line of equality. The velocities are considered to be in agreement if they are within 5\,\kms; agreement for the low and high reliability clumps is $\sim$80\,per\,cent.
     }
		\label{fig:velo_offsets}
	
\end{figure}

As can be seen from the velocities shown in Fig.\,\ref{fig:velocity_examples} there are some significant differences in the velocities allocated here  (i.e., Sect.\,\ref{sec:vel}) and by the \higal\ team (\citealt{mege2021}). In Figure\,\ref{fig:velo_offsets}, we show the differences in the velocities for the 2\,725 clumps that measurements are available in both catalogues.

Differences of $<5$\,\kms\ are easily accounted for by the lower resolution of the survey data used by the \higal\ team to assign velocities. Differences of more than 5\,\kms, however, are likely to be due to the presence of multiple velocity components detected in the low-resolution spectra and the incorrect choice of velocity previously assigned to the clumps. Agreement is found for 2\,093 clumps (i.e. within 5\,\kms), corresponding to 78\,per\,cent of the entire sample. The agreement for high- and low-reliability sources is 76\,per\,cent (1232/1612) and 77\,per\,cent (861/1113), respectively. The large number of clumps where the velocities disagree (632) is somewhat surprising given that \citet{mege2021} used a similar methodology to assign velocities, including looking for morphological agreement in cases where multiple components were detected. 

We investigated these differences by first checking our calibration steps and by comparing the integrated emission in $\ell b$ and $\ell v$ with the maps of \citet{dame2001} but these checks did not reveal any issues with our data. The next step was to look at differences in the CO data used to assign velocities by the two studies. The \higal\ team used data from the NANTEN $^{12}$CO\,(1-0) survey in this part of the Galaxy, which has a spectral resolution of 1\,\kms\ and angular resolution of $\sim$156\,arcsec; however, a grid spacing of 4\,arcmin was used for $|b| < 5\degr$ and so this survey is undersampled. OGHReS, therefore, has significantly improved angular and spectral resolutions compared to the NANTEN survey. In addition, OGHReS uses a higher$-J$ CO transition and is more sensitive to denser gas than is traced by CO\,(1-0).

\begin{figure}
	\centering
        \includegraphics[width=0.45\textwidth]{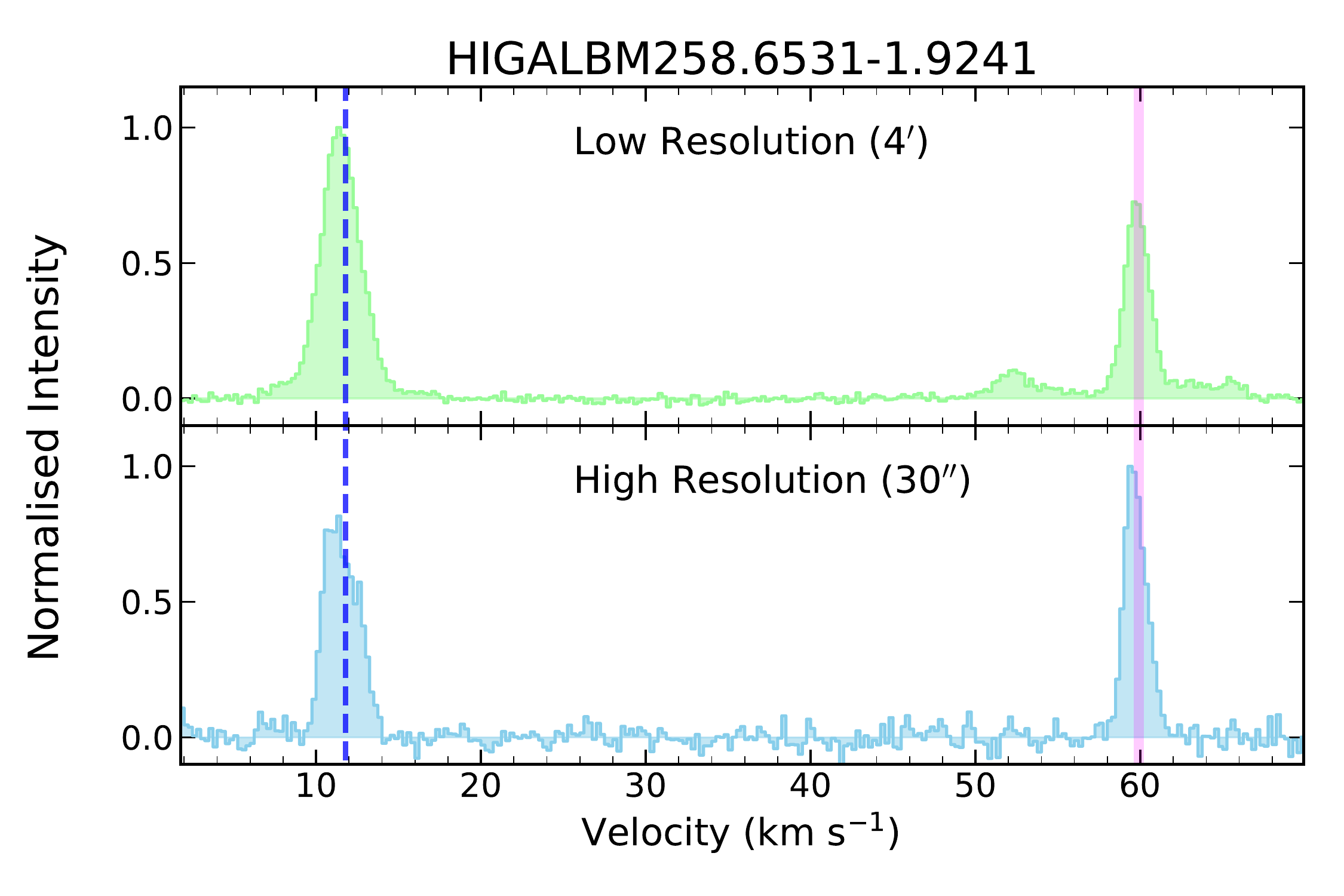}
        \includegraphics[width=0.45\textwidth]{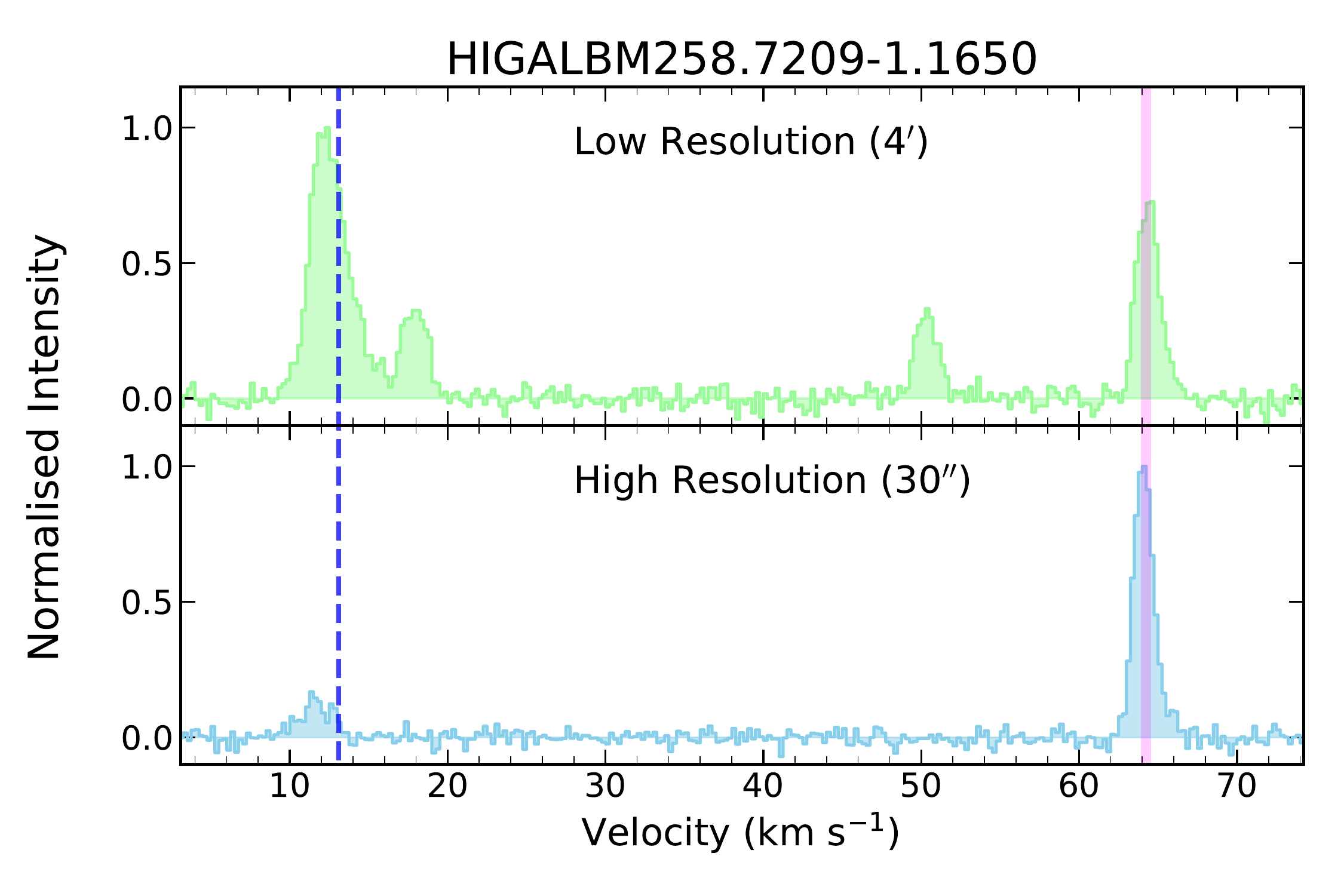}
        \includegraphics[width=0.45\textwidth]{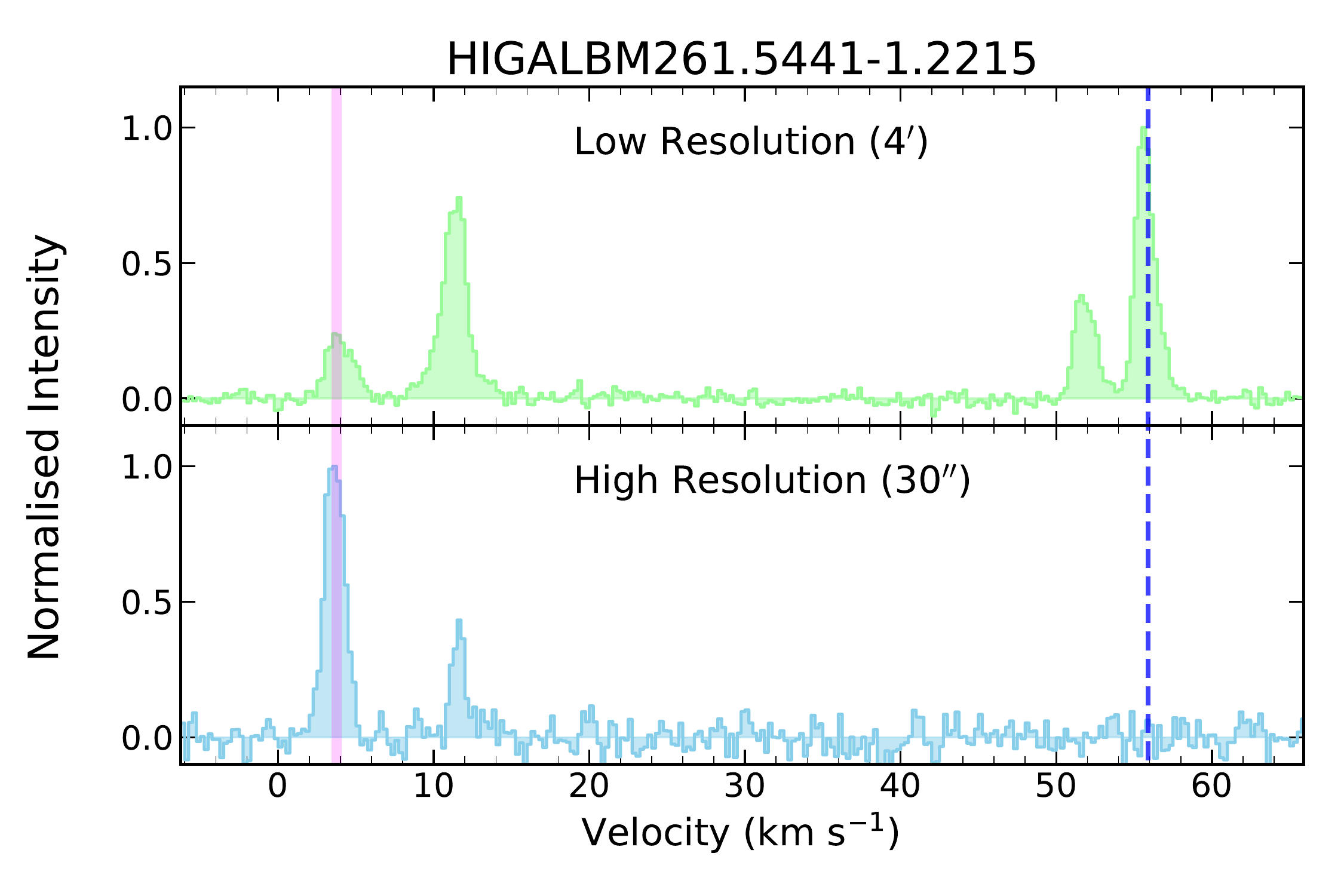}
    	\caption{$^{12}$CO\,(2-1) line emission toward selected \higal\ clumps. The low-resolution spectra have been produced by integrating the emission within a 2\,arcmin radius of the \higal\ clump to simulate the emission seen in the NANTEN survey. The high-resolution spectra is the native OGHReS used to determine the velocity in this study. All of the spectra have been normalised to the highest intensity component in each spectra.   The blue vertical dashed line shows the velocity allocated by the \higal\ team (\citealt{mege2021}) and the thick vertical magenta line shows the velocity allocated from the OGHReS data in this paper. }
		\label{fig:hilow_spectra}
	
\end{figure}

The NANTEN data are not publicly available and so making a direct comparison between the spectra used by the \higal\ team with the spectra we have is not possible, however, we can compare high-resolution OGHReS spectra with low-resolution OGHReS spectra extracted using a 4-arcmin beam to simulate the emission seen in the NANTEN data. In Figure\,\ref{fig:hilow_spectra}, we show some examples of the high- and low-resolution OGHReS spectra. The upper panel shows a similar profile, indicating that the emission in the low-resolution spectrum is emitted from a similar region to the high-resolution spectrum. However, the low-resolution spectra shown in the middle and lower panels of Fig.\,\ref{fig:hilow_spectra} are very different from the high-resolution spectra, revealing strong emission features at velocities that have no significant emission in the high-resolution spectra. The low-resolution spectra tend to have more components, reflecting the more complex CO emission structure present within a 4-arcmin beam. Furthermore, the emission from the compact clumps detected by \higal\ is affected by beam dilution and emission from more diffuse gas becomes more prominent.

The spectra shown in Fig.\,\ref{fig:hilow_spectra} demonstrate the impact that the angular resolution has on the number and strength of emission components and explains why there can be differences in the velocities allocated here and those allocated by the \higal\ team. To evaluate the impact of the improved resolution and tracer, we have looked at the correlation of the velocities for clumps where \oghres\ only detects a single CO component (i.e., where there is no ambiguity in the source velocity). Of the clumps where the velocities disagree, we have detected only a single CO component towards 374 clumps and blended components towards a further 59 clumps (i.e., all components within 5\,\kms\ of each other) and so there is no ambiguity in our velocity allocation; this corresponds to $\sim$70\,per\,cent of the disagreements. We have used integrated $^{12}$CO maps to allocate a velocity to another 187 sources and so are confident in these allocations. The remaining 12 sources have been assigned the velocity of the strongest component, as it was more than twice the intensity of any other components detected; these velocity allocations are the least certain. These checks confirm that the OGHReS velocity allocations are reliable for approximately 98\,per\,cent of the sample (620/632) where we have found disagreements with the \higal\ velocities.

The improved resolution and sensitivity of the OGHReS survey data provide reliable velocities to 3\,412 \higal\ clumps, including corrected velocities to 632 clumps and velocities for a further 687 \higal\ clumps where a velocity has not previously been determined. We have therefore adopted these velocities for the whole \higal\ sample of clumps. The \higal\ catalogue provides velocities to 97 clumps, which are not allocated a velocity from the OGHReS data, either due to no CO emission components being detected (51 objects) or detection of multiple components and where a reliable velocity cannot be identified. Since we are unable to confirm the \higal\ velocities independently, we do not include them in the analysis that follows.


\subsection{Galactic Distribution}

\begin{figure*}
	\centering

  \includegraphics[width=0.99\linewidth]{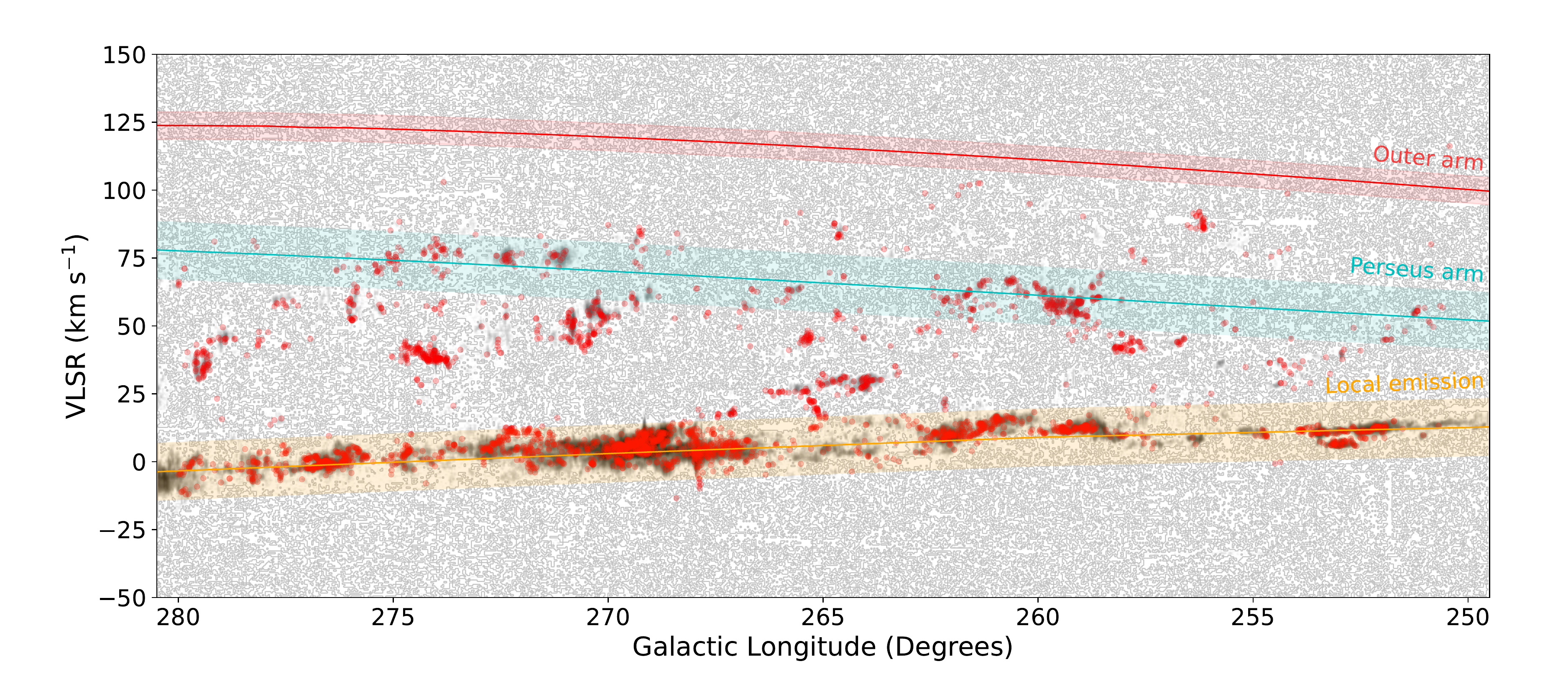}

    	\caption{Integrated $^{12}$CO longitude-velocity map of a 30 square degree contiguous region of OGHReS ($249.5\degree \leq \ell \leq 280.5\degree$; grey scale). The position and updated velocities of the \higal\ clumps are shown as filled red circles. The Perseus and Outer arm models are shown in cyan and red, respectively, and are taken from \citet{reid2014}. The local arm is also shown as a yellow band. The emission has been integrated over one degree in latitude (i.e. $-2\degree \leq b \leq -1\degree$). } 
     
		\label{fig:long_vlsr_dist}
	
\end{figure*}

In Figure\,\ref{fig:long_vlsr_dist}, we show the longitude-velocity distribution of all \higal\ clumps in the OGHReS longitude range investigated in this paper. The  distribution of dense compact clumps traced by \higal\ mirrors that of the more diffuse molecular gas. There is a reasonable correlation between the dense clumps and the local and Perseus arms. These are also, however, a significant number of clumps located between these two arms. The majority of these interarm clumps are associated with filamentary structures, many of them spur like, as identified by \citet{colombo2021} using the full OGHReS data.

We also note that there is almost no correlation between the dense clumps and the loci of the Outer arm. A significant number of clumps are located in the inter-arm region between the Perseus and Outer arm. This population, however, falls off towards the Outer arm with only two clumps being in its vicinity. The Outer arm is known to dip below $b = -2$\degr\  in this part of the Galaxy due to the warp (\citealt{nagayama2011}), which is not covered by OGHReS.

\setlength{\tabcolsep}{3pt}

\begin{table*}


\begin{center}\caption{Updated \higal\ clump parameters. The distances have been determined from the OGHReS velocities determined here and the gas-to-dust ratio ($\gamma$) from \citet{giannetti2017}. The rest of the properties have been scaled from those given in the \higal\ catalogue using the updated distances and gas-to-dust ratios. The integers given in the Evolution type column correspond to unbound = 0, starless = 1 and protostellar = 2.}
\label{tbl:derived_clump_para}
\begin{minipage}{\linewidth}
\small
\begin{tabular}{lc....c.....}
\hline \hline
  \multicolumn{1}{c}{\higal}&  
  \multicolumn{1}{c}{Reliability} &
  \multicolumn{1}{c}{Evolution}&
  \multicolumn{1}{c}{\vlsr} & 
  \multicolumn{1}{c}{Distance} &
  \multicolumn{1}{c}{R$_{\rm{gc}}$}&
  \multicolumn{1}{c}{$\gamma$} &
  \multicolumn{1}{c}{Diameter} & 
  \multicolumn{1}{c}{Log[$M_{\rm{clump}}$]} & 
  \multicolumn{1}{c}{$\Sigma_{\rm gas}$} &  
  \multicolumn{1}{c}{Log[$L_{\rm{bol}}$]}	&
  \multicolumn{1}{c}{Log[$L/M$]} \\

    \multicolumn{1}{c}{name }&  
    \multicolumn{1}{c}{}&
    \multicolumn{1}{c}{type} & 
    \multicolumn{1}{c}{(\kms)}  &
    \multicolumn{1}{c}{(kpc)} &
    \multicolumn{1}{c}{(kpc)}&
     \multicolumn{1}{c}{} &
    \multicolumn{1}{c}{(pc)}&
    \multicolumn{1}{c}{(\msun)} &
    \multicolumn{1}{c}{(g\,cm$^{-2}$)} &
    \multicolumn{1}{c}{(\lsun)} &
    \multicolumn{1}{c}{(\lsun/\msun)}   \\
    
\hline

HIGALBM250.6483-1.9895	&	high	&	1	&	57.4	&	5.1	&	10.9	&	245.5	&	0.34	&	1.994	&	0.228	&	0.602	&	-1.391	\\
HIGALBM250.6871-1.1550	&	high	&	1	&	13.3	&	1.3	&	8.7	&	156.1	&	0.14	&	1.459	&	0.375	&	-0.235	&	-1.694	\\
HIGALBM250.8094-1.9433	&	high	&	0	&	49.7	&	4.4	&	10.4	&	223.0	&	0.66	&	1.668	&	0.028	&	0.478	&	-1.189	\\
HIGALBM250.8631-1.6673	&	high	&	2	&	80.1	&	7.3	&	12.6	&	343.7	&	0.33	&	1.564	&	0.108	&	1.873	&	0.309	\\
HIGALBM251.0410-1.0367	&	high	&	1	&	9.8	&	1.0	&	8.5	&	151.8	&	0.12	&	1.705	&	0.904	&	-0.009	&	-1.714	\\
HIGALBM251.1581-1.9913	&	high	&	1	&	56.8	&	5.0	&	10.9	&	242.6	&	0.51	&	2.388	&	0.243	&	1.255	&	-1.133	\\
HIGALBM251.1729-1.9826	&	high	&	2	&	55.3	&	4.9	&	10.8	&	238.2	&	0.69	&	2.697	&	0.275	&	2.010	&	-0.687	\\
HIGALBM251.1801-1.9766	&	high	&	2	&	55.5	&	4.9	&	10.8	&	238.7	&	0.95	&	2.939	&	0.258	&	2.739	&	-0.200	\\
HIGALBM251.1919-1.9733	&	high	&	2	&	55.5	&	4.9	&	10.8	&	239.2	&	0.90	&	2.821	&	0.220	&	3.267	&	0.445	\\
HIGALBM251.1931-1.9758	&	high	&	1	&	55.5	&	4.9	&	10.8	&	239.2	&	0.43	&	3.263	&	2.640	&	1.577	&	-1.685	\\

\hline\\
\end{tabular}\\
Notes: Only a small portion of the data is provided here. The full table is available in electronic form at the CDS via anonymous ftp to cdsarc.u-strasbg.fr (130.79.125.5) or via http://cdsweb.u-strasbg.fr/cgi-bin/qcat?J/MNRAS/.
\end{minipage}

\end{center}
\end{table*}
\setlength{\tabcolsep}{6pt}

\subsection{Kinematic Distances}

\begin{figure}
	\centering
        \includegraphics[width=0.45\textwidth]{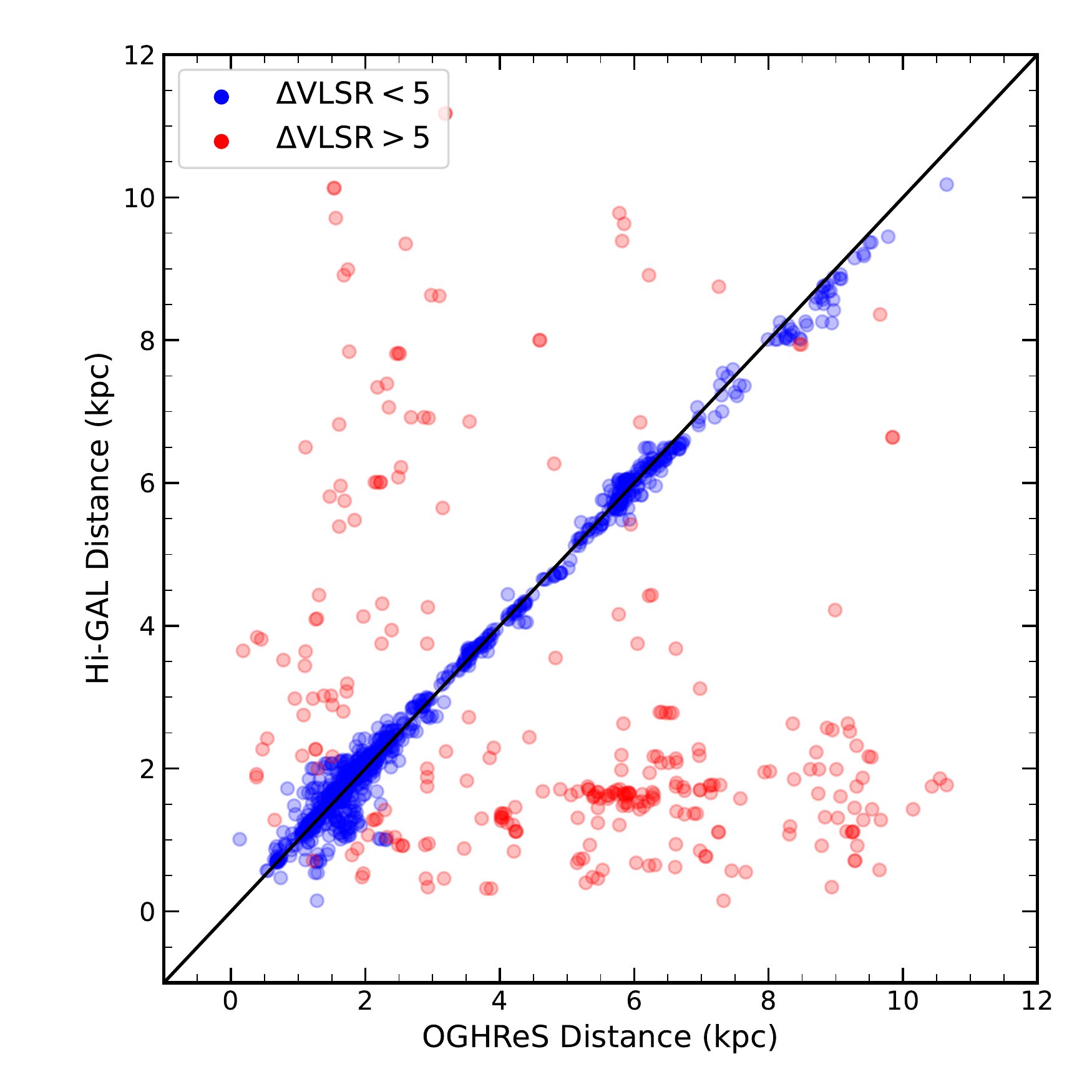}
    
    	\caption{Comparison of the heliocentric distances determined from the \higal\ survey and the OGHReS where the velocities have been assigned by the two surveys and the assigned distances are kinematic in nature. The line of equality is shown in black. }
		\label{fig:distance_comparison}
	
\end{figure}

\begin{figure*}
	\centering

        \includegraphics[width=0.45\textwidth]{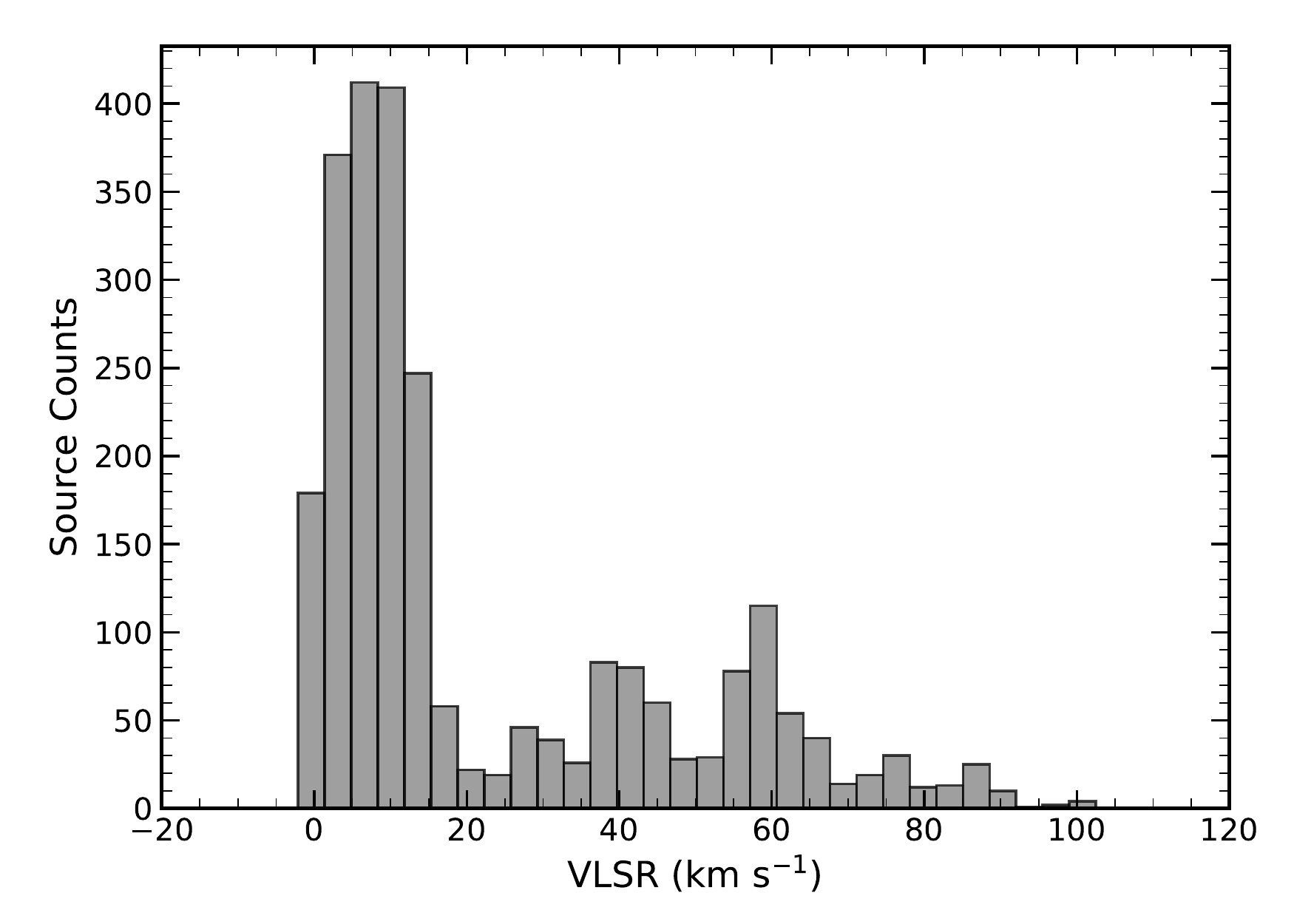}
         \includegraphics[width=0.45\textwidth]{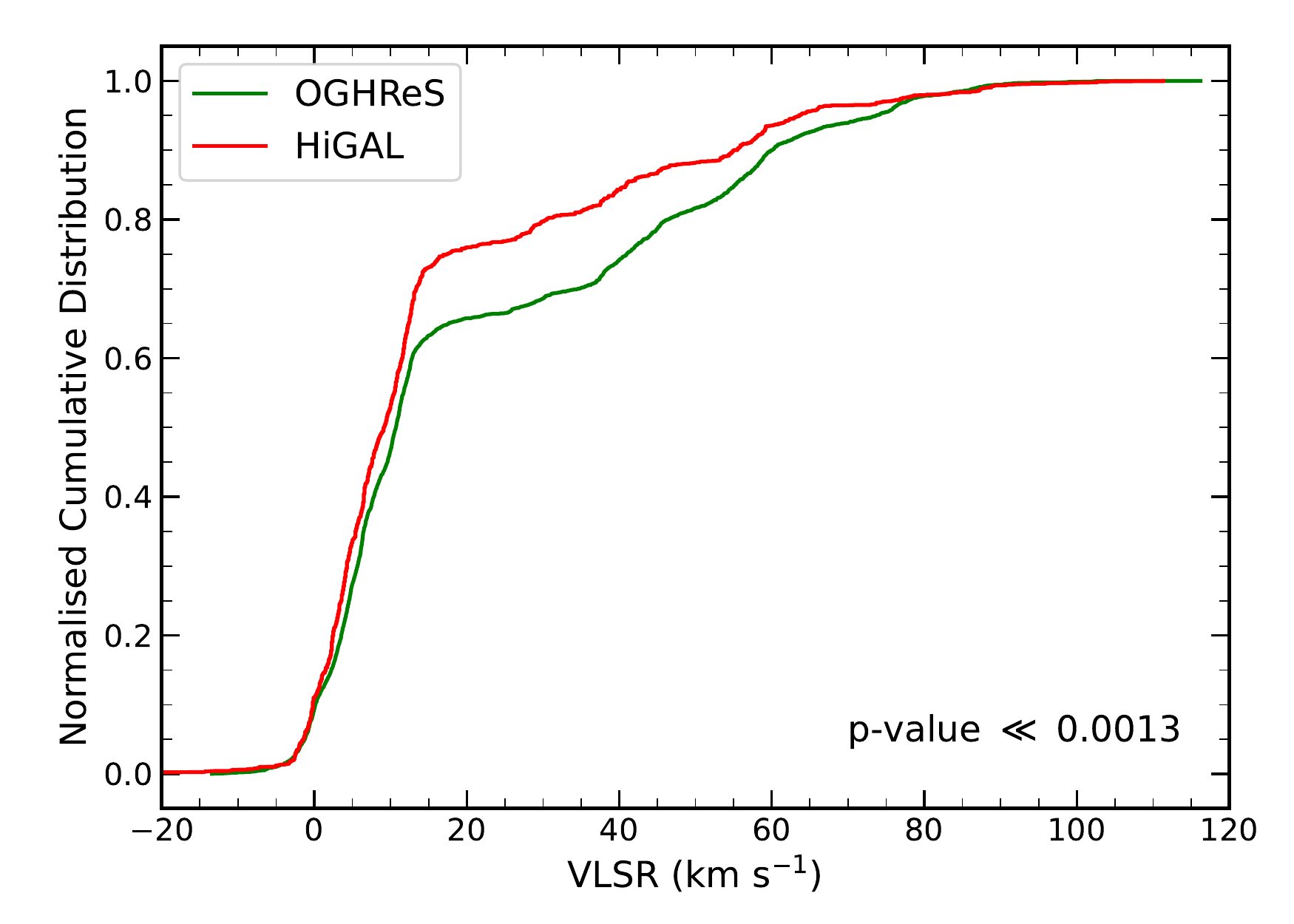}\\
         \includegraphics[width=0.45\textwidth]{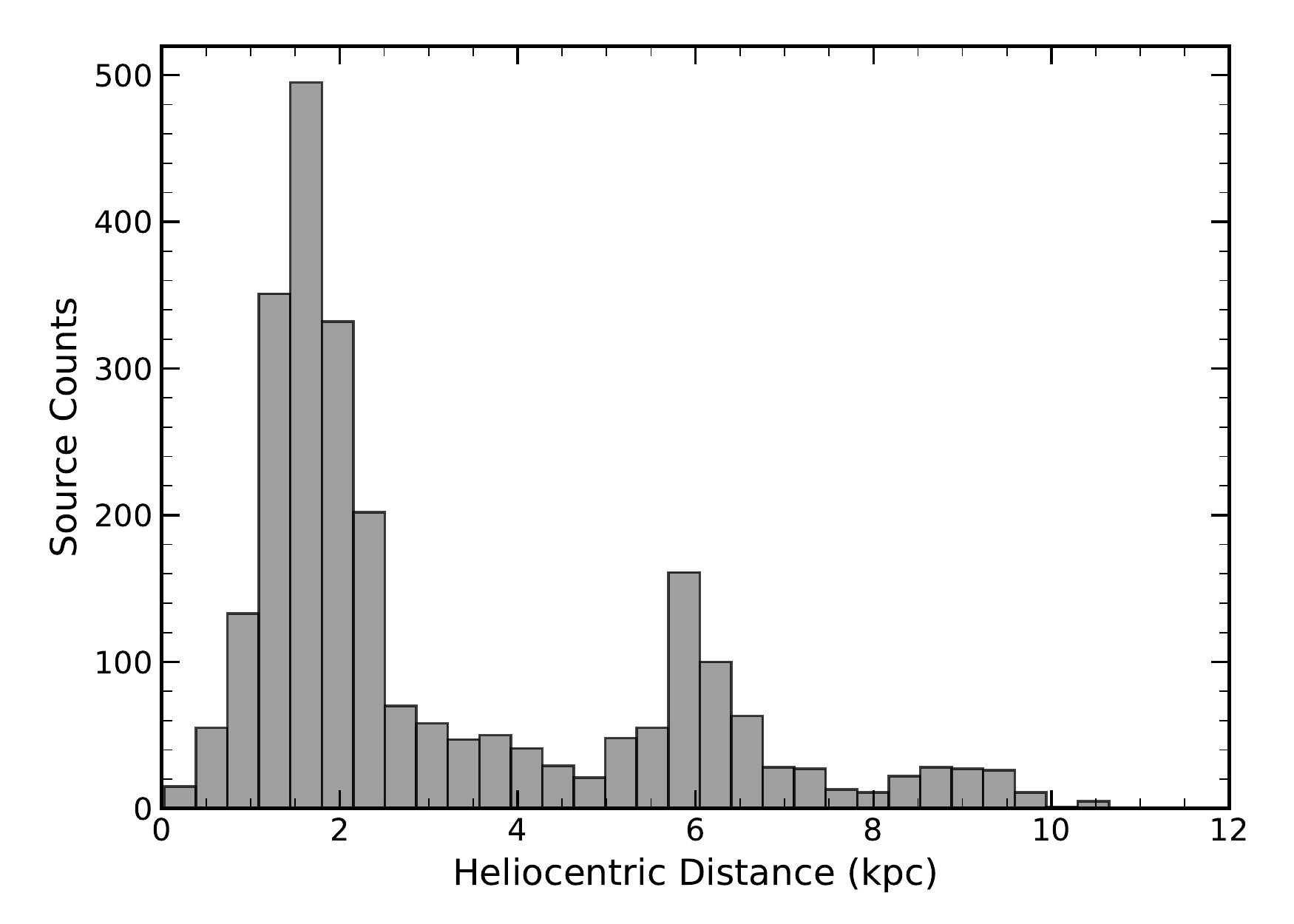}
         \includegraphics[width=0.45\textwidth]{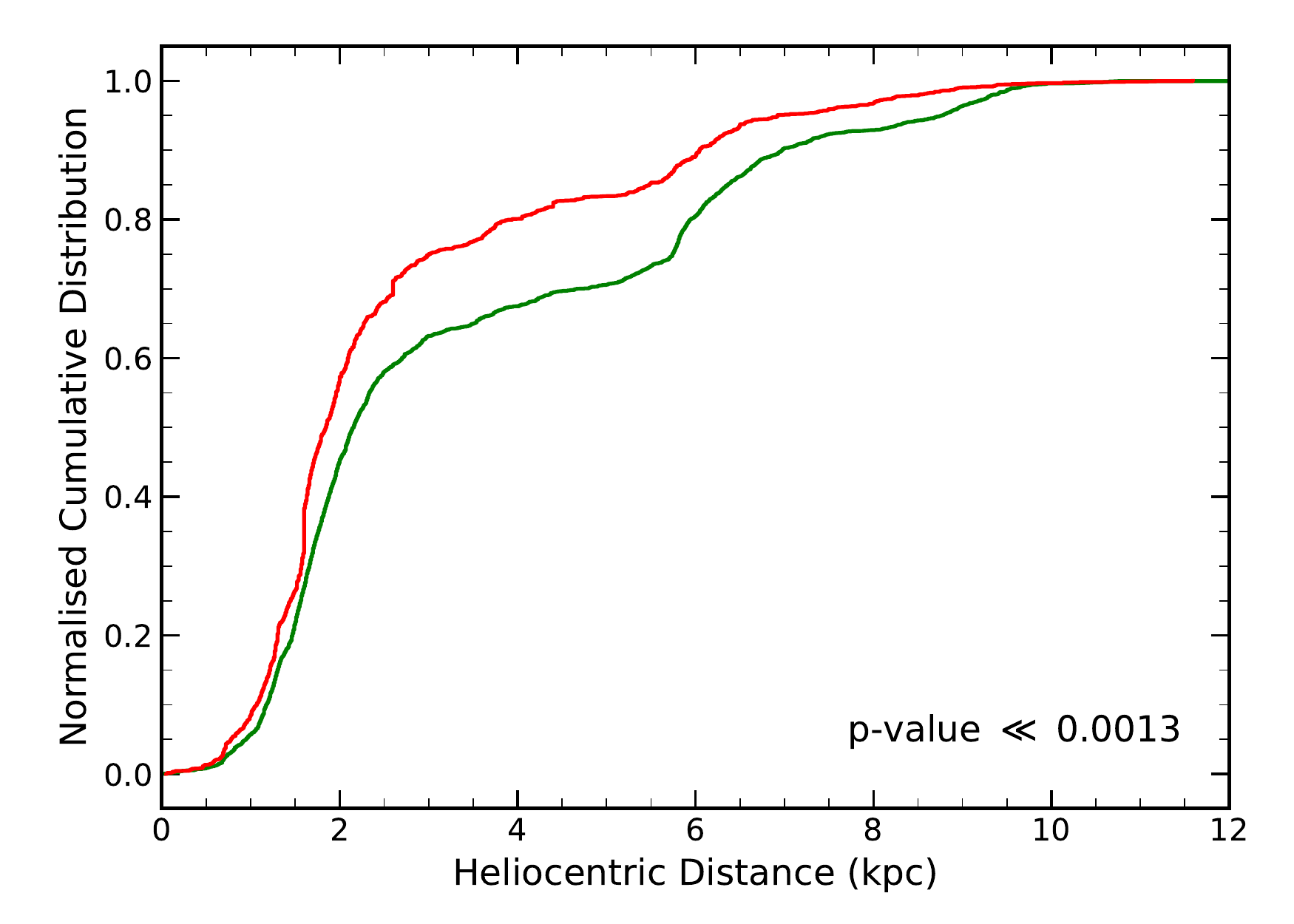}\\
         \includegraphics[width=0.45\textwidth]{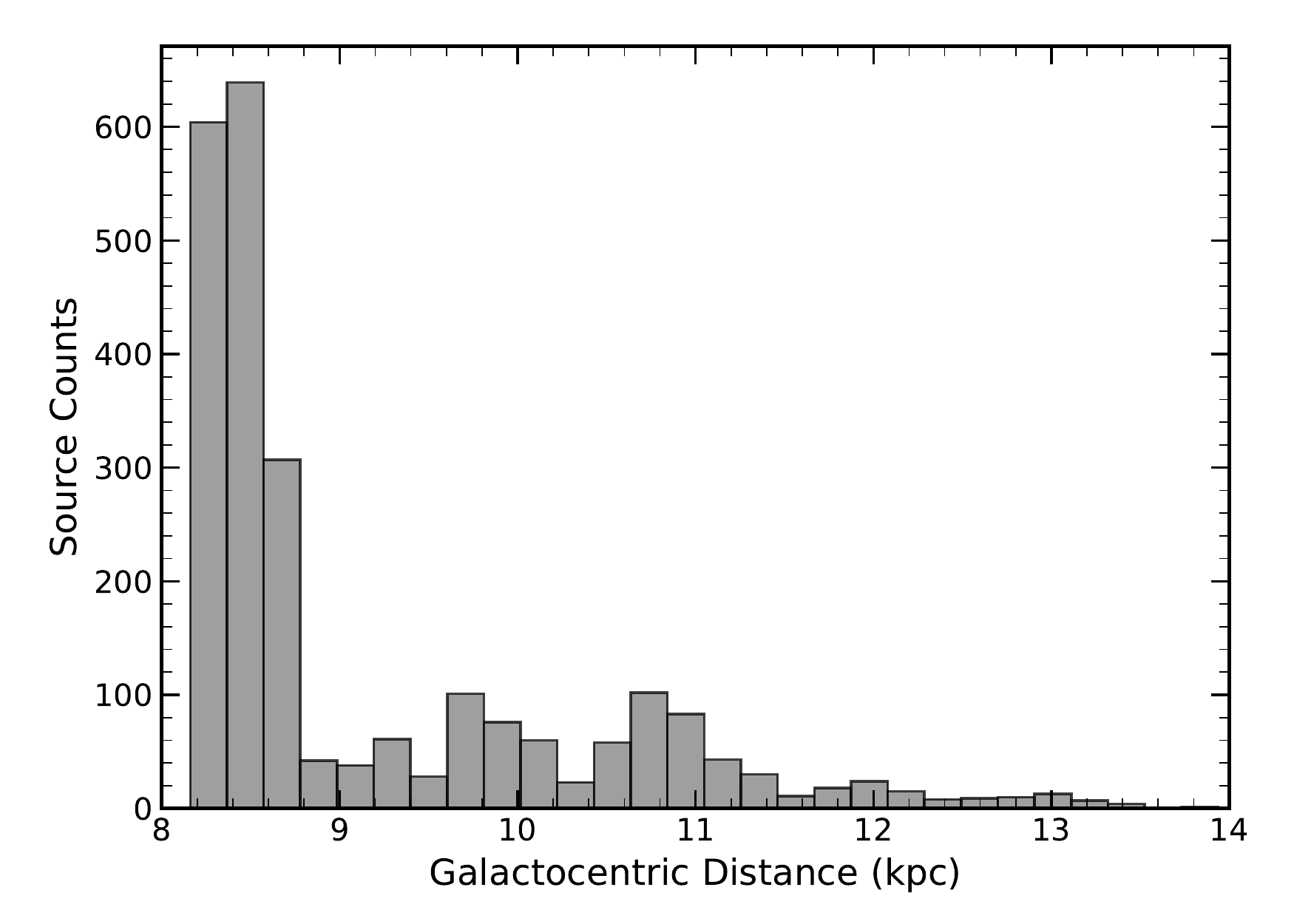}
         \includegraphics[width=0.45\textwidth]{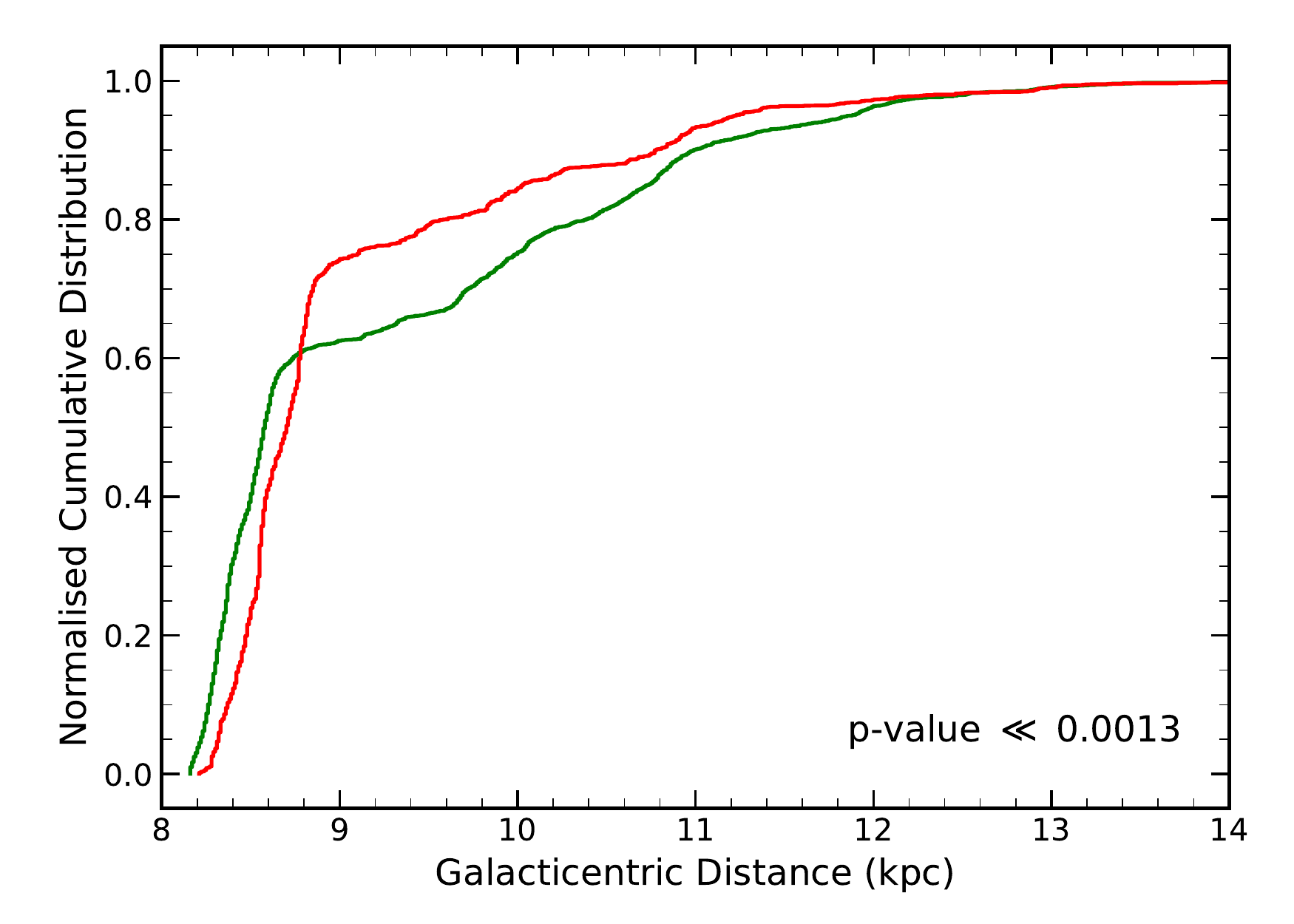}
        
    	\caption{Velocity and distance distributions based on the OGHReS and \higal\ surveys. The left panels show histograms of updated velocity, heliocentric and Galactocentric distance distributions. The right panels show the corresponding cumulative distribution functions of the \higal\ measurements (red) and the updated values reported here (green). The $p$-values for the results of the \ks\ tests on the two sets of measurements are given in the lower right corner.}
		\label{fig:distance_velocity_hist}

\end{figure*}

By combining these radial velocities with a Galactic rotation curve, we can determine kinematic distances and Galactocentric distances for the clumps. The distances in the \higal\ catalogue were determined by \citet{mege2021} using the rotation curve of \citet[][$R_0 =  8.34$\,kpc , $\theta_0 = 240$\,\kms]{russeil2017}. To be consistent with the previous OGHReS paper (i.e. \citealt{colombo2021}) and the SEDIGISM survey (\citealt{schuller2021,cabral2021}), however,  we use the \citet{brand1993} rotation curve ($R_0 = 8.15$\,kpc and $\theta_0 =240$\,\kms; see \citealt{colombo2021} for discussion of these parameters). One significant advantage of using a rotation curve to assign distances in the outer Galaxy is there is no kinematic distance ambiguity, which affects all sources located within the \SC\ (i.e. $R_{\rm gc} < 8.15$\,kpc) and so the distances are more reliable. This provides heliocentric and Galactocentric distances to 3\,200 clumps.

In Figure\,\ref{fig:distance_comparison}, we compare the distances obtained from the two different rotation curves using the velocities derived by \higal\ and OGHReS data for the same sources. In this plot, we indicate clumps where the velocities agree within 5\,\kms\ and those that differ by more than 5\,\kms\ with different colours to emphasise the impact on the derived distances. This plot reveals a very strong agreement between the distances using the two curves for those where the velocities are in good agreement and, hence, that the choice of rotation model is not crucial to this work.

In the left panels of Figure\,\ref{fig:distance_velocity_hist}, we show the velocity and the heliocentric and Galactocentric distributions of the \higal\ sources derived from the velocities determined in the previous section. All three distributions have a similar profile with a strong peak at low values and two weaker peak at higher values. The strongest peak is associated with the local arm with the weaker structures associated with the sources in the interarm region and the Perseus arm (cf. Fig.\,\ref{fig:long_vlsr_dist}).  In the right panels of Fig.\,\ref{fig:distance_velocity_hist}, we show the corresponding cumulative distribution plots of both the refined values determined in this section with the OGHReS data and those reported in the  \higal\ catalogue (\citealt{elia2021}). These plots reveal that the updated values are significantly different and this difference is confirmed by a \ks\,test ($p$-values $\ll 0.0013$).

\subsection{Physical properties}

We have used the updated and more complete set of distances to re-calculate the distance-dependent physical properties of the \higal\ clumps (i.e. luminosity, mass, and radius) and to revisit some of the comparisons presented in \citet{elia2021}. The \higal\ team uses a gas-to-dust ratio of 100 for all sources. While this approach is likely to be reasonable for the inner part of the Galactic disc, it is unlikely to provide reliable masses and surface density measurements over the whole disc. When updating the physical parameters to take account of the new distances, we have also taken the opportunity to include the variation in the gas-to-dust ratio $\gamma$ as a function of Galactocentric distance \citep{giannetti2017}:

\begin{equation}
    {\rm log}(\gamma) = \left(0.087\left[^{+0.045}_{-0.025}\right]\pm 0.007 \right)R_{\rm gc} +  \left(1.44\left[^{+0.21}_{-0.45}\right]\pm 0.03 \right),
\end{equation}

\noindent where $R_{\rm gc}$ is the Galactocentric distance in kpc. The systematic uncertainties are given in the square brackets. This prescription gives a value of $\gamma$ between 130 and 145 at the distance of the Sun, which is in very good agreement with the local value of 136 (see \citealt{giannetti2017} for a detailed discussion). Although the luminosity to mass ratio and surface density are distance independent quantities, they are affected by the gas-to-dust variations and so we also update these accordingly. Although \citet{elia2021} used a constant gas-to-dust ratio, they did speculate on the impact that using a variation in this ratio may have and so incorporating it here is a logical progression of their work. In Table\,\ref{tbl:derived_clump_para}, we provide the updated values for the \higal\ clumps.

\begin{figure}
	\centering
       
        \includegraphics[width=0.45\textwidth]{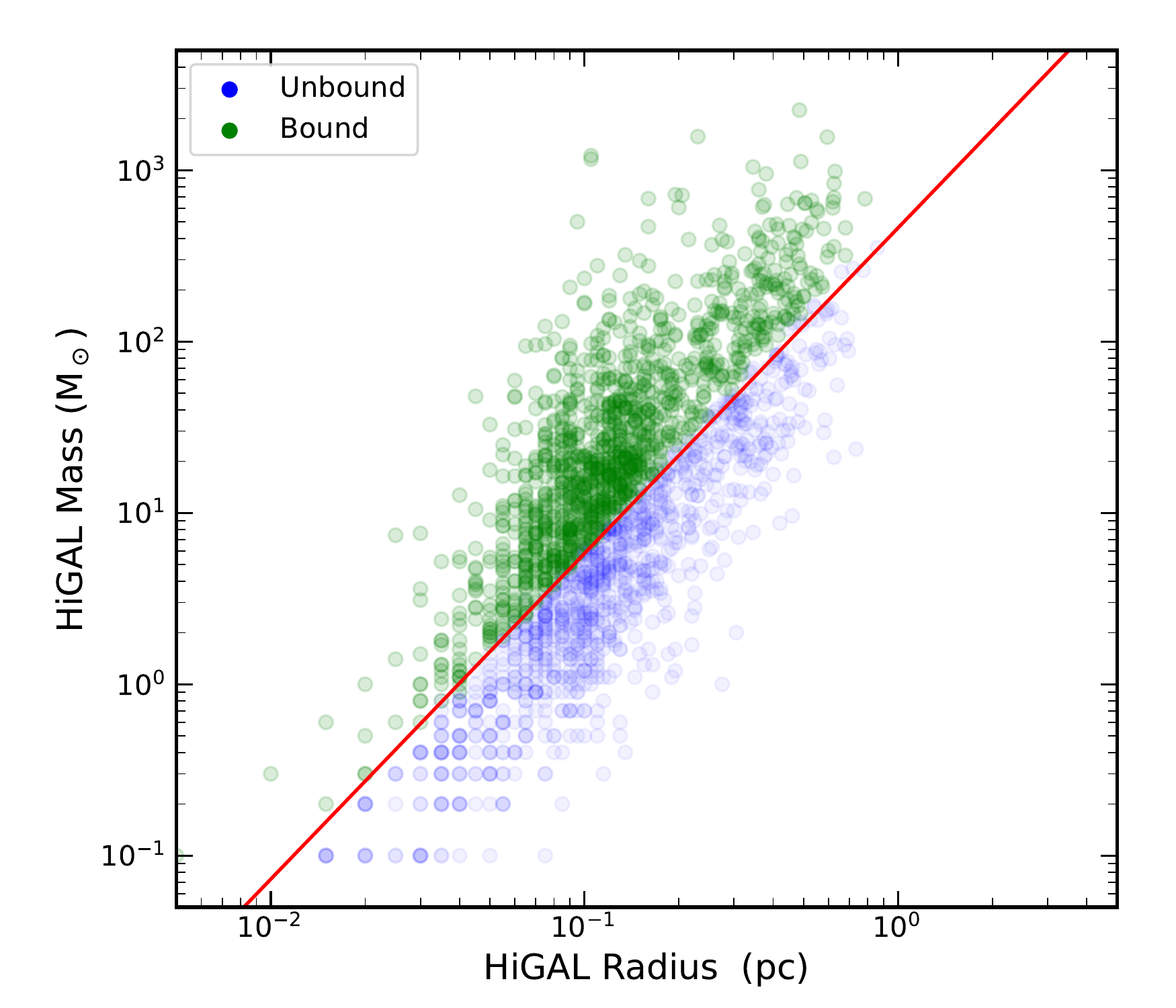}
        \includegraphics[width=0.45\textwidth]{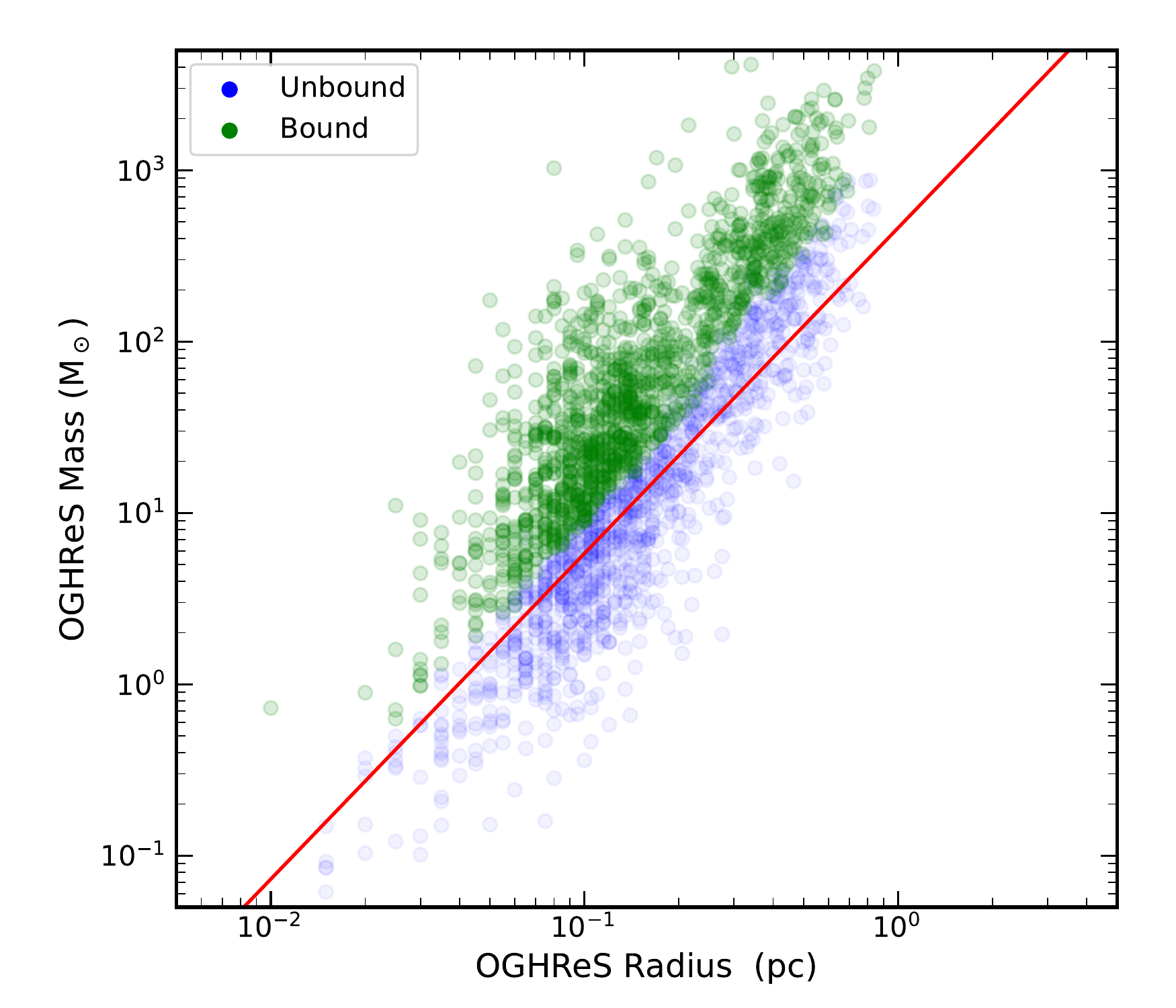}

    	\caption{Distinguishing between bound and unbound clumps. The red line shows Larson's third law (i.e. $M(r) = 460$\,\msun\,$(r/{\rm pc})^{1.9}$), which is used to distinguish between bound and unbound clumps; clumps above this threshold are considered bound while clumps below it are considered unbound. The upper panel shows the classification of clumps in the OGHReS region using the masses and sizes given in the \higal\ catalogue \citep{elia2021}. The lower panel shows the impact on the \higal\ classification with the recalculated masses and sizes in this paper. All unbound clumps located above the mass-size threshold have been reclassified as bound.}
		\label{fig:larson_3rd_law}

\end{figure}

Updating the distances, masses and sizes also affects the classification of the clumps. This is because mass is proportional to distance squared but size is proportional to distance i.e., mass increases faster with distance than the radius and consequently nearby clumps that appear to be unbound can become bound if found to be a farther away. In the vast majority of cases where we have modified the distance given in the \higal\ catalogue, it has resulted in an increase in the distance determined. In addition to the changes in distance, the change in the gas-to-dust ratio also results in higher masses in the outer Galaxy, which also works to change the status from unbound to bound for many clumps. In Figure\,\ref{fig:larson_3rd_law}, we show how the changes to the physical parameters impact the classification of clumps unbound to bound. These changes result in 739 unbound clumps being reclassified (415 to bound and 324 to unclassified as a distance is not available) and two bound clumps being reclassified as being unbound. This reduces the unbound population by a factor of 2. The change of distance does not affect the protostellar clump population, as these are classified by their association with a 70-\mum\ point source.

\section{Galactic trends revisited}

\begin{figure*}
    \centering
   \includegraphics[width = 0.49\textwidth, trim=0 0 0 0]{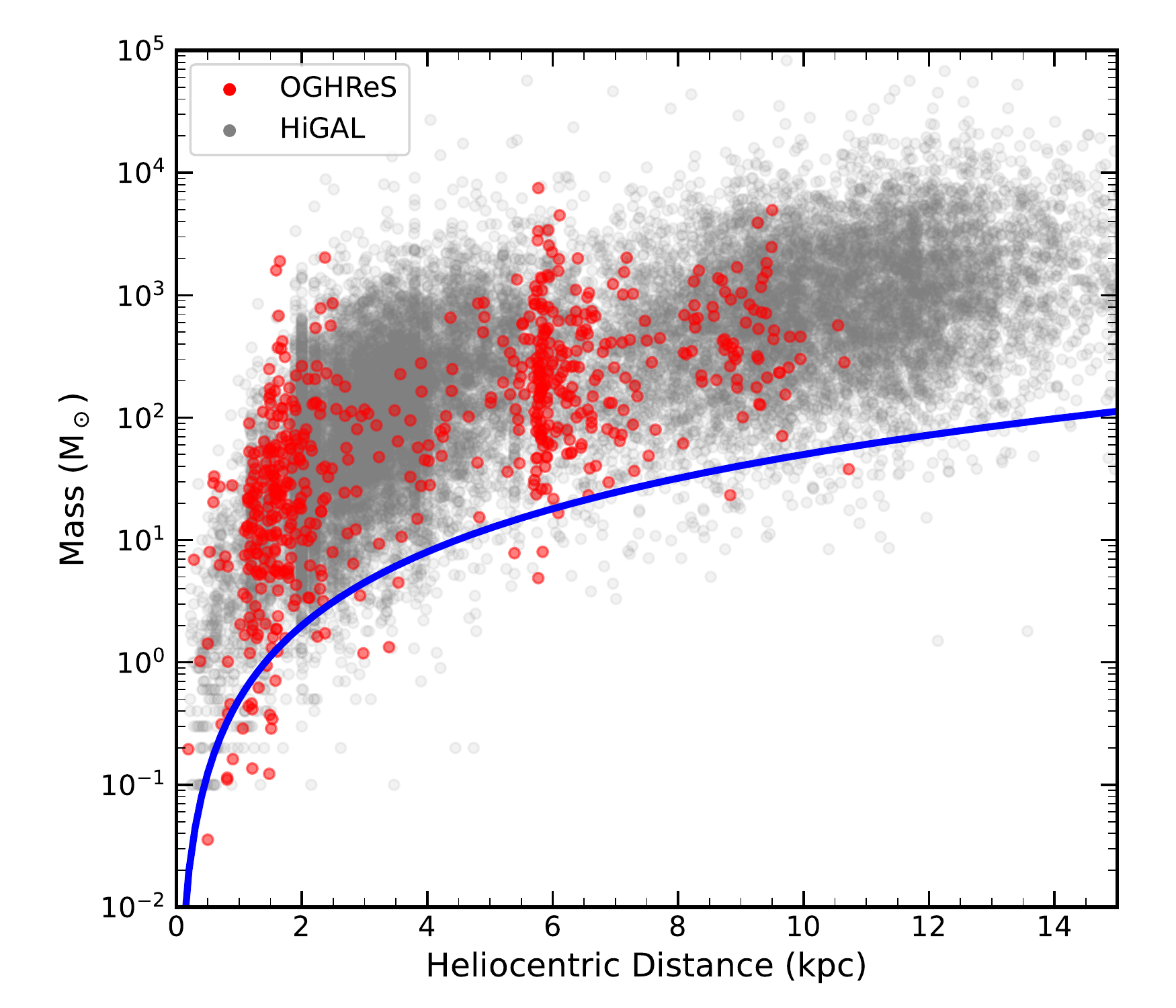}
    \includegraphics[width = 0.49\textwidth, trim=0 0 0 0]{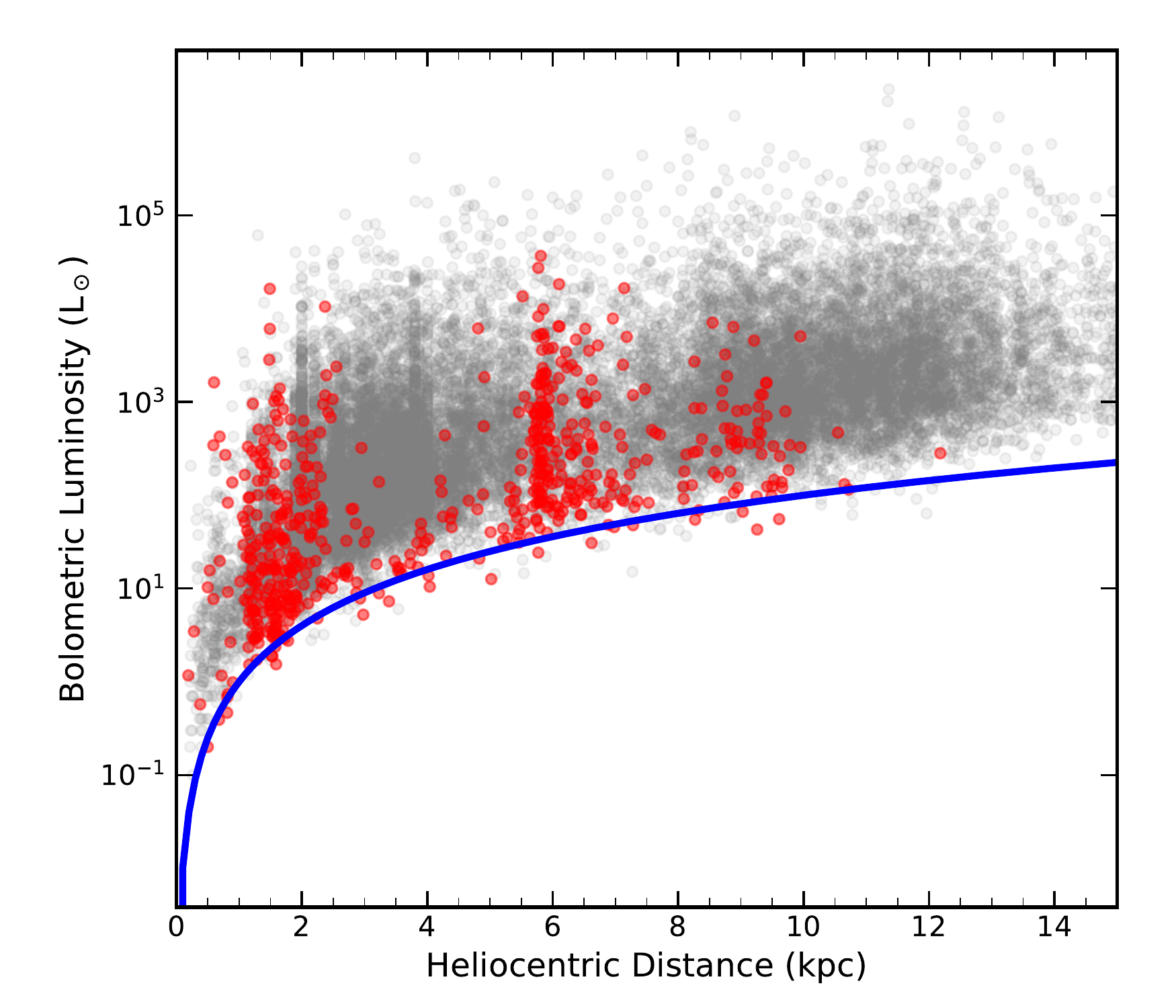}

    \caption{Distribution of clump mass and bolometric luminosity as a function of heliocentric distance are shown in the left and right panels respectively. The \higal\ sources located within the inner Galaxy are shown in grey and the clump located in the OGHReS region  are shown in red. The blue curves illustrate the impact of \higal\ flux sensitivity on the clump mass and luminosity detection limit as a function of distance (values are arbitrary).
    }
    \label{fig:dist_heliocentric}
\end{figure*}

\citet{elia2021} used the \higal\ high-reliability catalogue to investigate trends in the Galactic distribution of the clump mass, luminosity, surface density, and luminosity-to-mass ratio and reported significant differences between the inner and outer Galaxy. Given the improvements we have made with respect to completeness and the reliability of the clump velocities, distances, and distance dependent physical parameters, it is useful to revisit their analysis to evaluate the impact these updated values may have on their conclusions. It is important, however, to bear in mind that the data presented here represent a relatively small fraction of the \higal\ outer Galaxy sample (i.e. between $\ell = 250\degr$ and $280\degr$).

An issue with all flux-limited surveys is that the sensitivity to mass and luminosity decreases with distance. In Figure\,\ref{fig:dist_heliocentric}, we show the mass and luminosity distributions as a function of heliocentric distance of inner-Galaxy \higal\ (grey filled circles) and outer-Galaxy \higal\ clumps (red filled circles), as updated here. This clearly demonstrates the impact of the \higal\ flux sensitivity on these two parameters. We also note that the distribution of clumps as a function of distance for the inner- and outer-Galaxy samples are significantly different, with those in the inner Galaxy being more evenly distributed while the outer Galaxy sources are more localised between 1-2\,kpc, 5-7\,kpc and $\sim 9$\,kpc. To minimise the distance bias, and differences in the distributions, when comparing the properties of the inner and outer Galaxy samples we define a distance limited sample of clumps between 5 and 7\,kpc.

\begin{table*}
\centering
\caption{Median values for physical parameters for the high-reliability \higal\ catalogue \citep{elia2021}, subdivided by evolutionary class and inner/outer Galaxy location.}
\label{tbl:inner_vs_outer}
\begin{tabular}{lccc c ccc}
\hline
 & \multicolumn{3}{c}{Inner Galaxy} & &\multicolumn{3}{c}{Outer Galaxy} \\
\cline{2-4} \cline{6-8}
 &\multicolumn{1}{c}{Unbound}& \multicolumn{1}{c}{Bound} &\multicolumn{1}{c}{Protostellar} & &  \multicolumn{1}{c}{Unbound} & \multicolumn{1}{c}{Bound} & \multicolumn{1}{c}{Protostellar} \\
\hline
 \multicolumn{8}{c}{All} \\
 \hline 
$M_{\rm clump}\, [{\rm M}_\odot] $ & 27.7 & 226 & 218 &&  6.35 & 51.7 & 77.7 \\
$L_\mathrm{bol} [{\rm L_\odot}]$ & 33.5 & 56.3 & 553 && 2.60 & 4.36 & 89.1 \\
$L_\mathrm{bol}/M_{\rm clump}\, [{\rm L_\odot/M_\odot}]$ & 1.17 & 0.25 & 2.76 && 0.42 & 0.08 & 1.35 \\
$\Sigma$\, [g\,cm$^{-2}$]  & 0.03 & 0.14 & 0.22 && 0.03 & 0.13 & 0.23 \\
\hline 
 \multicolumn{8}{c}{ $D < 8.15$\,kpc}\\
\hline 
$M_{\rm clump}\, [{\rm M}_\odot] $ & 16.7 & 109 & 94.6 &&  6.26 & 47.6 & 72.8 \\
$L_\mathrm{bol} [{\rm L_\odot}]$ & 20.5 & 24.9 & 189 && 2.37 & 3.94 & 86.4 \\
$L_\mathrm{bol}/M_{\rm clump}\, [{\rm L_\odot/M_\odot}]$ & 1.05 & 0.22 & 2.46 && 0.40 & 0.08 & 1.42 \\
$\Sigma$\, [g\,cm$^{-2}$]  & 0.03 & 0.15 & 0.26 && 0.03 & 0.13 & 0.23 \\
\hline
 \multicolumn{8}{c}{ $5\,{\rm kpc} <D < 7\,{\rm kpc}$}\\
\hline 
$M_{\rm clump}\, [{\rm M}_\odot] $ & 55.9 & 279 & 264 &&  58.3 & 311 & 208 \\
$L_\mathrm{bol} [{\rm L_\odot}]$ & 47.6 & 66.2 & 561 && 22.7 & 20.5 & 277 \\
$L_\mathrm{bol}/M_{\rm clump}\, [{\rm L_\odot/M_\odot}]$ & 0.80 & 0.23 & 2.53 && 0.40 & 0.07 & 1.45 \\

$\Sigma$\, [g\,cm$^{-2}$]  & 0.03 & 0.13 & 0.22 && 0.03 & 0.16 & 0.20 \\
\hline
\end{tabular}

\end{table*}

In Table\,\ref{tbl:inner_vs_outer}, we present the median values for clump mass, luminosity, surface density, and luminosity-to-mass ratio for the three different clump evolutionary stages defined by \citet{elia2021}. These are split into inner- and outer-Galaxy samples and only include clumps in the high-reliability catalogue with a distance.\footnote{This is slightly different to \citet{elia2021}, as they used a statistical method for determining whether a source without a distance was in the inner or outer Galaxy.  The inclusion of these sources, however, has a significant impact on the median values determined and represents a potential bias we prefer to avoid.}  When comparing the inner- and outer-Galaxy samples, we have also applied the modified gas-to-dust ratio to the inner Galaxy clumps. We further separate these into three different distance samples. In the upper part of the table we include the whole inner- and outer-Galaxy sample to facilitate comparison with the values provided by \citet{elia2021}. The masses and luminosities are a factor of 4 and 10 times higher in the inner Galaxy than in the outer Galaxy. In the middle section of Table\,\ref{tbl:inner_vs_outer}, we show the values for the sample of clumps within 8.15\,kpc of the Sun. Restricting the sample to the near side of the Galaxy results in a more similar distribution in clump mass (the difference in the unbound, bound and protostellar clumps masses are within a factor of 3, 2 and 1.5, respectively) in the inner and outer Galaxy and the difference in luminosities is reduced to a factor of two. 

In the lower section of Table\,\ref{tbl:inner_vs_outer}, we give the values for the distance-limited sample. The values themselves are somewhat arbitrary as they depend on the distance range used, but the comparison should be a reliable guide to the difference between the physical properties of inner- and outer-Galaxy clumps.  We find that the masses in the inner and outer Galaxy are similar (within $\sim$20\,per\,cent), with the inner-Galaxy clumps being a little more massive, and the luminosity of inner-Galaxy clumps being a factor of 2--3 more luminous. The large differences in the clump masses and luminosities in the inner and outer Galaxy seen when compiling values from the whole sample are, therefore, the result of observational biases. 

The median masses of the unbound clumps are significantly lower than those of sources in the other two evolutionary stages. The masses of bound and protostellar clumps are broadly similar to each other in both the inner and outer Galaxy.  With respect to the median luminosities, we find that starless clumps (unbound and bound) are similar within a factor of two, as one might expect, given that their luminosity represents thermal emission from the dense gas, but there is a sharp increase for the protostellar clumps (factor of $\sim10$).

Looking at the $L/M$-ratio we find it to be a factor of 2-3 higher in the inner Galaxy than in the outer Galaxy for all three evolutionary stages. There is a clear jump in the $L/M$-ratio between the pre-stellar (bound) and protostellar clumps in both the inner and outer Galaxy, as one would expect. The $L/M$-ratio for the unbound clumps are up to 5 times larger than for the bound clumps. The unbound clumps, however, are either pressure confined or transient and their luminosity is likely to be dominated by the interstellar radiation field, which is able to penetrate more deeply than for bound clumps. Their higher median $L/M$-ratios are probably a result of these environmental factors, and strictly speaking, it only makes sense to consider the $L/M$-ratio and corresponding evolutionary tracks for  gravitationally bound clumps (see also \citealt{elia2021} for a more detailed discussion).

\begin{figure*}
    \centering
    \includegraphics[width = 0.49\textwidth, trim=0 0 0 0]{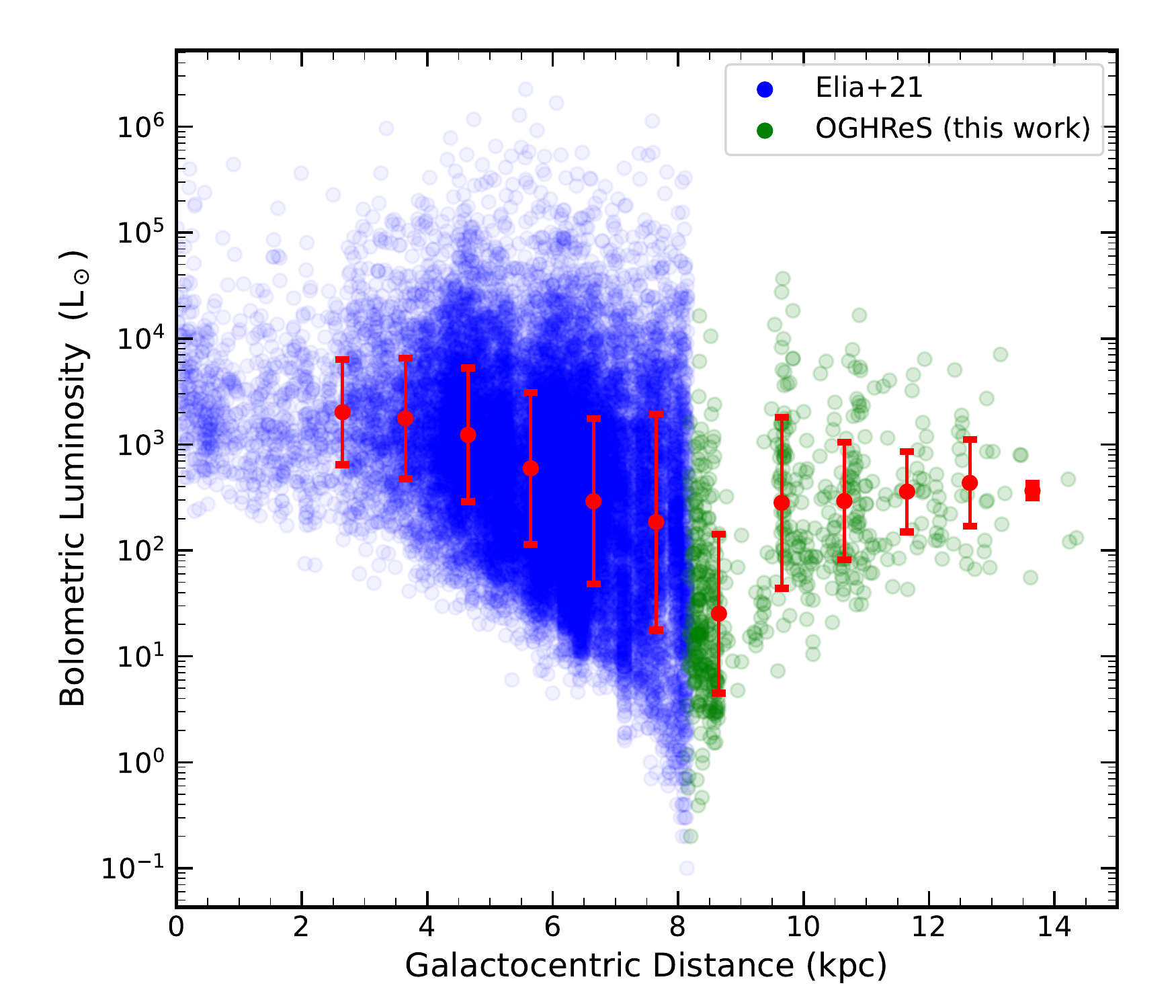}
    \includegraphics[width = 0.49\textwidth, trim=0 0 0 0]{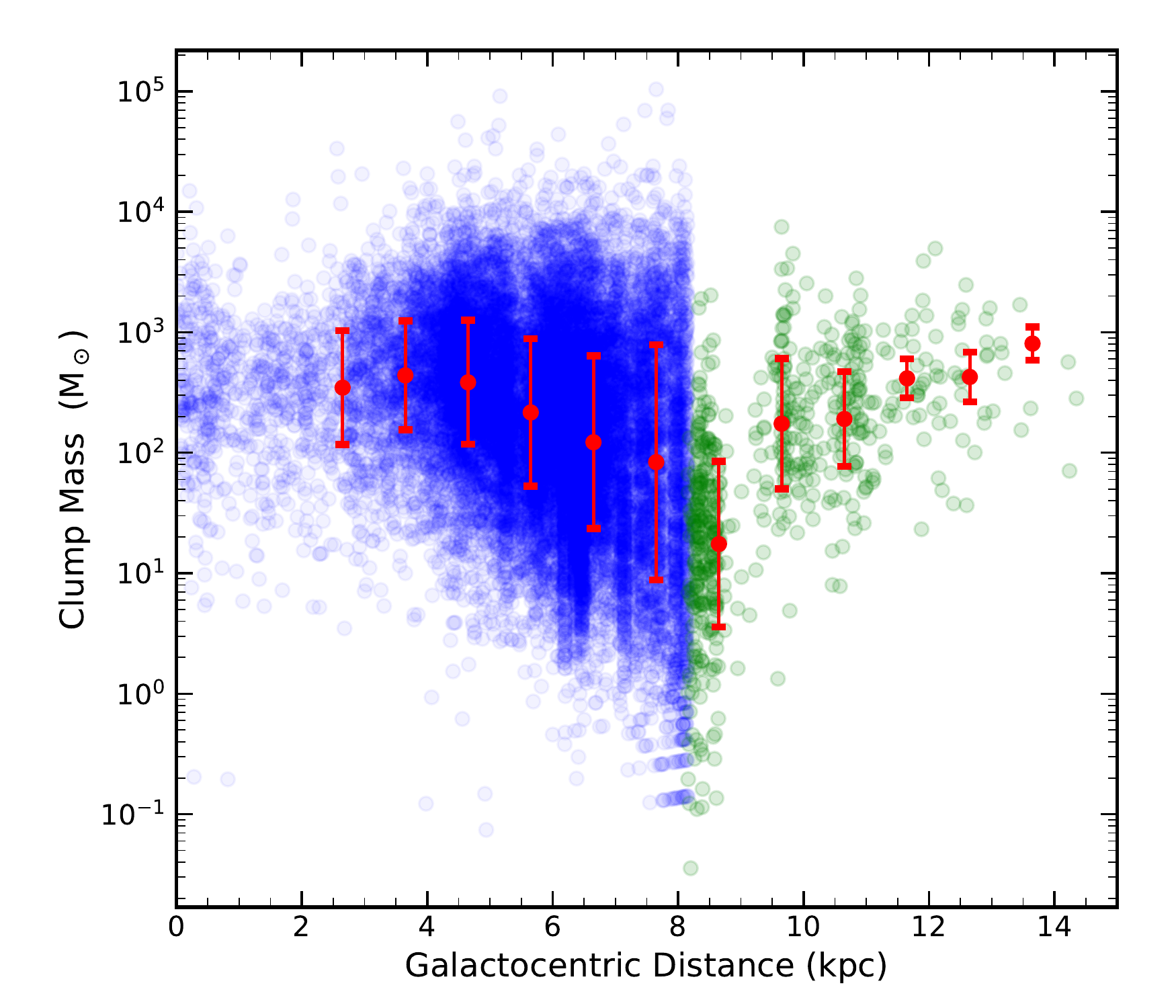}
    
    \includegraphics[width = 0.49\textwidth, trim=0 0 0 0]{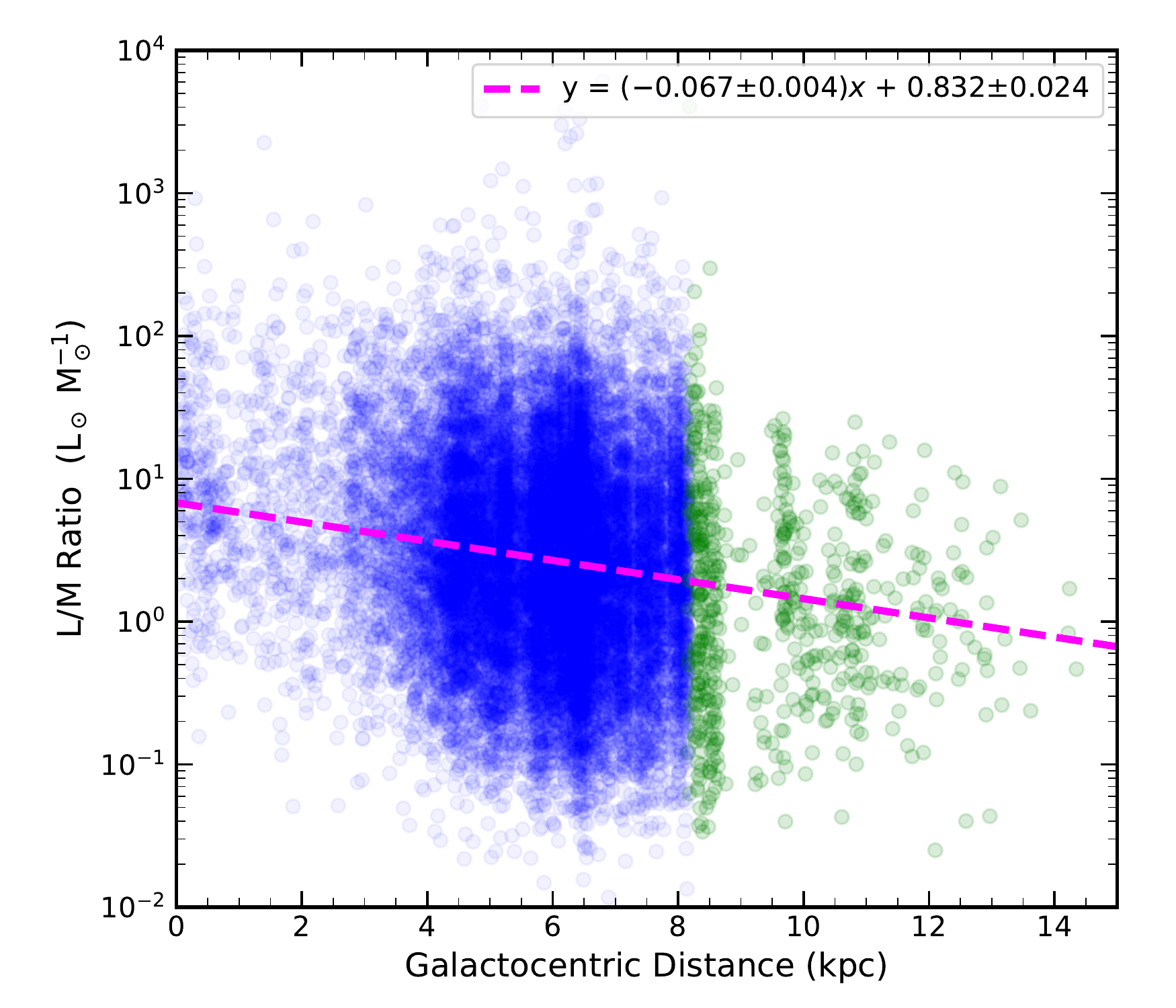}
    \includegraphics[width = 0.49\textwidth, trim=0 0 0 0]{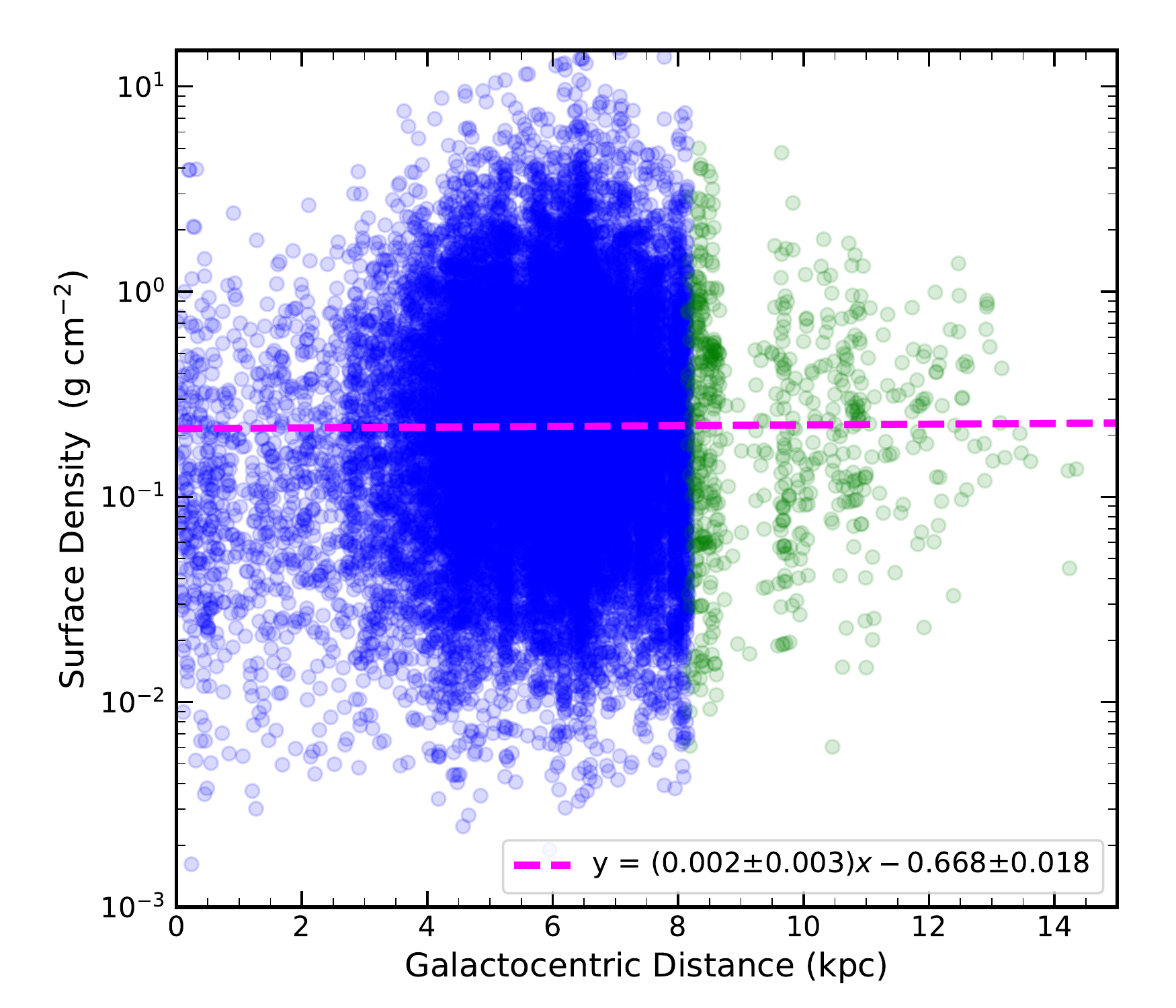}
    \caption{Physical parameters of \higal\ protostellar clumps as a function of Galactocentric distance. The clump properties have been updated with a gas-to-dust ratio relationship determined by \citet{giannetti2017} and the distances of the outer Galaxy clumps in the OGHReS region have been updated with the results presented in this work. The red circles and error bars shown in the upper panels are the lognormal  mean and standard deviation calculated from the clumps within each 1\,kpc wide bins between 2 and 14\,kpc. The dashed magenta line plotted in the lower panels shows the results of  linear least squares fits to the plotted data and the fit parameters are given in the respective insets. The fits has been made to all data with Galactocentric distances larger than 3\,kpc and Galactic longitudes between 5\degr\ and 355\degr\ in order to exclude sources towards the Galactic centre region where distances are not considered to be reliable.
    }
    \label{fig:parameters_fn_rgc}
\end{figure*}

Comparing the surface density values for the three evolutionary samples for both the inner and outer Galaxy, we find that they show a similar trend of increasing median value with evolution. Comparing the values derived for the evolutionary samples, we again find them to be similar for the inner or outer Galaxy. This similarity is in contrast to the trend reported by \citet{elia2021} who found the surface densities for each evolutionary stage were lower in the outer Galaxy compared to clumps in the inner Galaxy. They did note that the decrease in the surface density as a function of Galactocentric distance was similar to that of the gas-to-dust ratio gradient reported by \citet{giannetti2017} and so, it is not surprising that having applied this correction we now find similar values for the surface density of clumps across the disc.

In Figure\,\ref{fig:parameters_fn_rgc}, we show how the parameters for the protostellar clumps change as a function of Galactocentric distance. In the top two panels of this figure we present the clump luminosities and masses. Both of these distributions show a minimum towards $\sim8$\,kpc. That Galactocentric distance corresponds to the location of the \SC, for which we have the best sensitivity to mass and luminosity, i.e. a lower bound set by completeness (as shown in Fig.\,\ref{fig:dist_heliocentric}). To aid with the comparison of these two parameters in the inner and outer Galaxy, we include the mean and standard deviation determined from the lognormal distribution of clumps within 1\,kpc wide bins between Galactocentric distances of 2\,kpc and 14\,kpc. It is clear from the upper left panel of Figure\,\ref{fig:parameters_fn_rgc} that the luminosity is lower in the outer Galaxy, which confirms the difference in mean values for the inner and outer Galaxy samples reported in Table\,\ref{tbl:inner_vs_outer}. The clump masses are broadly similar across the disc except for the first bin in the outer Galaxy, which is noticeably lower than the corresponding bin in the inner Galaxy (i.e. the 7.15 -- 8.15\,kpc bin). This outer Galaxy bin is dominated by local clumps, which also suppresses the median mass measurement for the outer Galaxy given in Table\,\ref{tbl:inner_vs_outer}. We note that this is also true for the luminosity distribution but this does not significantly affect the median value given in Table\,\ref{tbl:inner_vs_outer}.

The lack of correlation between clump mass and Galactocentric distance is a little surprising, given the difference in cloud mass between the inner and outer Galaxy reported by \citet{brand1995}. They found that between Galactocentric distances of 3\,kpc and 20\,kpc the mass of molecular clouds decreased by approximately a factor of 100, and that, even between the \SC\ and Galactocentric distances of $\sim$20\,kpc, the cloud masses decrease by an order of magnitude. An important distinction between their work and ours is that here we are focusing on star-forming clumps and so perhaps the conditions for star formation are the same across the disc (as discussed later in this section) but this does not preclude the structure and mass distribution of molecular clouds themselves being very different.  If this is the case, we might expect the number of clouds hosting dense, star-forming clumps to decrease as a function of Galactocentric distance. Exploring that hypothesis, however, is beyond the scope of this work but will be investigated in a future paper when the complete OGHReS cloud catalogue becomes available.

In the lower-left panel of Fig.\,\ref{fig:parameters_fn_rgc}, we show the $L/M$-ratio, which also reveals a decreasing trend with increasing Galactocentric distance. This decline is confirmed by a linear fit to the data (dashed line) that clearly shows the trend extends over the whole disc. This trend may indicate that clumps in the outer Galaxy are producing smaller clusters and/or lower-mass stars than the higher-mass star forming clumps found in the inner Galaxy. Indeed, the decrease in the $L/M$-ratio is opposite to the result reported by \citet{elia2021}, where they found a modest increase in the outer Galaxy for the protostellar clumps. This change in the $L/M$-ratio  between the two studies is almost entirely due to the difference in the gas-to-dust ratio, which increases the value of the denominator (mass) at larger Galactocentric distances.

This trend of decreasing $L/M$-ratio as a function of Galactocentric distance is supported by \citet{djordjevic2019} who used a combination of ATLASGAL clumps (\citealt{schuller2009, urquhart2014_atlas}) and the SCUBA-2 Ambitious Sky Survey (SASSy; \citealt{thompson2007}) to investigate the properties of embedded \hii\ regions identified from their radio continuum emission (e.g. \citealt{urquhart_radio_south, urquhart_radio_north, purcell2013, urquhart2013_cornish}). They investigated the $L/M$-ratio and the total Lyman photon flux luminosity to mass ratio and found a decrease in both of these ratios as a function of 
Galactocentric distance. The decrease in the Lyman photon flux indicates that there is less high-mass star formation per unit volume taking place in the outer Galaxy. This finding is consistent with the results of an earlier study by \citet{brand1991} that was based on analysis of the infrared properties of molecular clouds.

The $L/M$-ratio is often used as a proxy for the instantaneous star-formation efficiency (\citealt{molinari2008}, \citealt{urquhart2013_methanol}). In this context, the decreasing value we find can be interpreted as evidence of a steady decrease in the star-formation efficiency in the outer Galaxy. Another measure of efficiency is the star-formation fraction (SFF), which \citet{ragan2016} defined as the ratio of \higal\ 70-\mum-bright clumps to the whole population of \higal\ clumps. Those authors reported a decrease in the SFF as a function of Galactocentric distance over the inner part of the disc (i.e. 3--8\,kpc), which is consistent with our findings. \citet{elia2021} explored the SFF out to larger Galactocentric distances (see upper-left panel of their Figure 28) and find it increases sharply outside of the \SC. Their count, however, includes clumps seen on the other side of the inner disc, which are likely to include a larger fraction of protostellar objects, and so this nominal increase should be treated with caution. We have attempted to investigate whether the decrease in SFF continues in the OGHReS region but the sample is too small to draw any useful conclusions.

\begin{figure}
    \centering
    \includegraphics[width = 0.49\textwidth, trim=0 0 0 0]{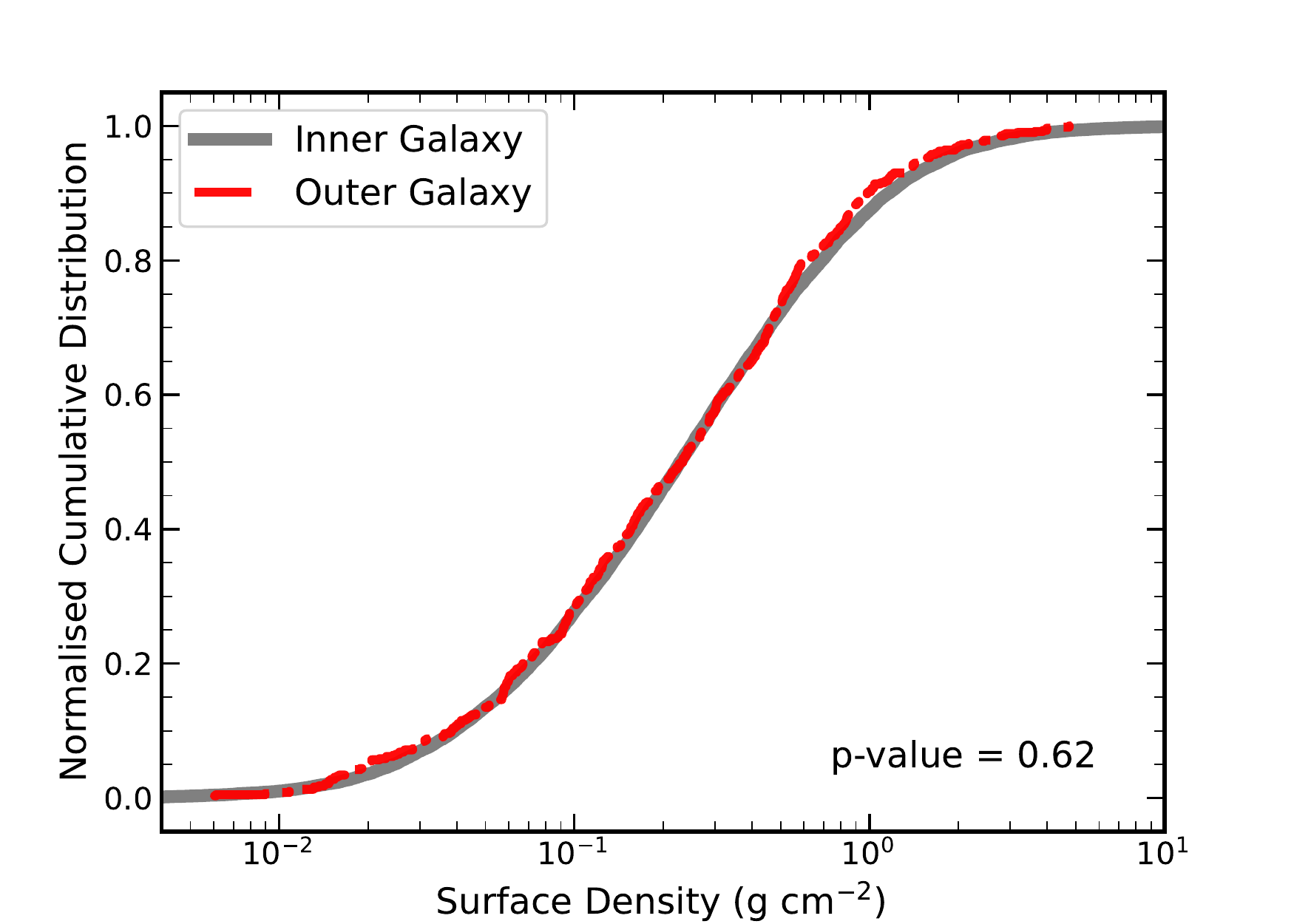}
  
    \caption{Cumulative distribution of the surface density of the inner- and outer-Galaxy protostellar clumps populations. The inner-Galaxy sample includes all protostellar clumps located within the disc (i.e., $3\,{\rm kpc} < R_{\rm gc} < 8.15\,{\rm kpc}$) while the outer-Galaxy sample includes all protostellar clumps located within the OGHReS region discussed here, where a distance has been determined and that lie outside the \SC\ (i.e. $R_{\rm gc} > 8.15\,{\rm kpc}$). }
    \label{fig:surface_density_inner_vs_outer}
\end{figure}

In the lower-right panel of Fig.\,\ref{fig:parameters_fn_rgc} we show the surface densities as a function of Galactocentric distance. We previously found  no significant difference between the surface densities for sources in the three evolutionary stages in the inner and outer Galaxy (see Table\,\ref{tbl:inner_vs_outer}) and this plot clearly shows  that the mean surface density is very similar across the whole Galaxy, however, we note that it varies over three orders of magnitude. The linear fit to the data between 3\,kpc and 14\,kpc reveals a very shallow increasing slope that is very close to zero when taking into account the uncertainty. In Figure\,\ref{fig:surface_density_inner_vs_outer}, we present the cumulative distribution for the inner and outer Galaxy protostellar clump populations. This figure robustly demonstrates the correlation between the two populations once the variation in the gas-to-dust ratio has been taken into account. The $p$-value is 0.52, which confirms the surface densities of inner and outer Galaxy protostellar clumps are statistically indistinguishable from each other.

This study focuses on a relatively small part of the outer Galaxy and so we need to be careful when drawing general conclusions. However, a similar conclusion was put forward by \citet{benedettini2021} from a study of dense clumps in the inner Galaxy (i.e. GRS \citealt{jackson2006} and SEDIGISM survey \citealt{schuller2017,schuller2021}) with the outer Galaxy (i.e. the Forgotten Quadrant Survey (FQS; \citealt{benedettini2020})).\footnote{The FQS covers $\ell = 220\degr - 240\degr$ and $b = 0\degr$ to $-$2.5\degr\ with $^{12}$CO and $^{13}$CO (1-0) and so is very complementary to the region presented in this paper.} They found the clump mass surface density to be very similar for the SEDIGISM and the FQS between 2\,kpc and 16\,kpc in Galactocentic distance. They report that the GRS clouds have a slightly higher mass surface density but suggested that to be the result of cloud crowding in the inner Galactic plane. The GRS has a slightly larger beam than SEDIGISM (46\,arcsec compared to 30\,arcsec) and used the $^{13}$CO (1-0) transition that is more easily excited and whose emission is subsequently more extended than the $^{13}$CO (2-1) used by SEDIGISM.  The results of these two studies leads us to conclude that the clump surface density is the similar across the Galaxy.

The similarity of the surface density for protostellar clumps indicates that the threshold for star formation across the disc is also likely to be similar. The lower $L/M$-ratio seen in the outer Galaxy, however, indicates that either the clumps  there are producing fewer stars or that they are producing relatively more lower-mass stars, i.e. star formation is either less efficient or the initial mass function (IMF) has a steeper slope in the outer Galaxy. A study by \citet{wouterloot1995} using infrared luminosities from IRAS measured the slope of the IMF in the outer Galaxy and found it to be indeed steeper than in the solar neighbourhood. Furthermore, a comparative study by \citet{brand2001} of a number of star-forming complexes in the far outer Galaxy found that, in the clumps in the outer-Galaxy clouds studied, gravity is the dominant force down to a lower mass than in local clouds. This may lead to the formation of more low-mass stars in the outer Galaxy than locally, which could explain the steeper IMF reported in the outer Galaxy \citep{brand2001}. 


Despite the similarity of the average surface density and mass of star-forming clumps in the inner and outer Galaxy, there is still a significant difference in the star-formation properties. This difference might be related to the higher temperatures found in the outer Galaxy (\citealt{elia2021}), the lower metallicity (e.g. \citealt{rudolph1997}), the diffuse Galactic interstellar radiation field (e.g. \citealt{bloemen1985}), or a combination or all three. Investigating the differences in the star formation in the inner and outer Galaxy is beyond the scope of this paper but is one of the main aims of OGHReS and will be explored in subsequent papers.

\section{Summary and conclusions}
\label{sect:summary}

We have used data from OGHReS, which is a new outer-Galaxy molecular-line survey, to refine the velocities, distances and physical properties assigned to a large sample of dense clumps identified by the \higal\ survey (\citealt{molinari2010a, elia2021}) located in the $\ell =  250\degr-280\degr$  and $-2\degr < b <-1\degr$ region. We have extracted $^{12}$CO\,(2-1) and $^{13}$CO\,(2-1) spectra towards 3\,584 clumps and determined the velocities of clumps along each line of sight. In cases where multiple components are detected towards a clump, making it difficult to assign a velocity unambiguously from the spectra alone, we created integrated emission maps of the different components and selected the velocity for which the CO-traced gas distribution best matches the position and morphology of the \higal\ clump. 

We have been able to allocate a reliable velocity for 3\,412 clumps (95\,per\,cent of the sample). Comparing these with the velocities given in the \higal\ catalogue (\citealt{elia2021}), we find good agreement for $\sim$80\,per\,cent of the sample (within 5\,\kms). We have investigated the reasons for the disagreements and have found that the lower angular resolution molecular-line data used previously to allocate velocities in the outer Galaxy by the \higal\ team (\citealt{mege2021}) is likely to have led to incorrect velocities being allocated. We consider velocities determined by the higher-resolution and sensitivity OGHReS data used here to be more reliable and therefore use these to calculate distances and update the physical properties of the \higal\ clumps accordingly.\\

\noindent Our main findings are:

\begin{itemize}

    \item  We have corrected the velocity for 632 clumps and provided velocities for 687 clumps to which no velocity had previously been allocated. We discard molecular-line velocities assigned by the  \higal\ team towards a further 97, of which 51 are non-detections in OGHReS and 69 are found to have multiple components for which we have been unable to determine a reliable velocity.  \\
    
    \item The new distances have been used in combination with a varying gas-to-dust ratio to produce an updated version of the \higal\ catalogue of dense clumps with reliable physical parameters. Reliable distances and physical properties are now available for 3\,200 outer Galaxy clumps ($\sim$90\,per\,cent of the \higal\ catalogue in this part of the Galaxy). This work represents a significant improvement.\\   

    \item Changes to the distances and masses in the outer Galaxy have resulted in the status of a 739 unbound clumps being changed (415 to bound and 324 to unclassified as a reliable distance is not available) and two clumps being reclassified from bound to unbound. This has reduced the unbound population of clumps by a factor of two in the region studied. \\

    \item Comparing the updated physical parameters for the \higal\ clumps in the inner and outer Galaxy, we find a clear trend for a decreasing luminosity-to-mass ratio with Galactocentric distance, with clumps in the outer Galaxy on average a factor of 2 lower. This trend suggests either that the star formation efficiency is significantly lower in the outer Galaxy or that a higher fraction of lower-mass stars are being formed compared to the inner Galaxy.\\

    \item We find that the surface density and mass of protostellar clumps is similar on average across the whole Galaxy indicating that the conditions required for star formation in clumps is also similar.\\
    
\end{itemize}

 This work demonstrates the utility of the OGHReS survey data and the improvement that its sensitivity and resolution has provided over what was previously available. Furthermore, given that this subsample represents only a third of the full coverage of the survey, it is already clear that it will have a lasting impact on our understanding of star formation in the outer Galaxy. 

\section*{Acknowledgements}

We thank the referee for their positive and timely response and for their comments and suggestions that have helped improve the clarity and readability of this work. AK acknowledges support from the Polish National Agency for Academic Exchange grant No. BPN/BEK/2021/1/00319/DEC/1. MF acknowledges support from the Polish National Science Centre via the grant
UMO-2022/47/D/ST9/00419. This research made use of \texttt{ASTROPY}\footnote{\href{https://www.astropy.org}{https://www.astropy.org}}, a community-developed core \texttt{PYTHON} package for Astronomy (\citealt{astropy2013,astropy2018}),  \texttt{NumPy} \citep{numpy}\footnote{\href{https://numpy.org}{https://numpy.org}}, \texttt{SciPy} \citep{scipy}\footnote{\href{https://www.scipy.org}{https://www.scipy.org}}, and \texttt{Matplotlib} \citep{matplotlib}\footnote{\href{https://matplotlib.org/}{https://matplotlib.org/}}.  This document was prepared using the Overleaf web application, which can be found at www.overleaf.com.
This research has made use of the VizieR catalogue, operated at CDS, Strasbourg, France. 

\section*{Data Availability}

The \higal\ catalogue used for this work can be found at \href{http://vialactea.iaps.inaf.it/vialactea/public/}{http://vialactea.iaps.inaf.it/vialactea/public/} and the OGHReS data used here can be found on CDS.

\bibliographystyle{mnras}
\bibliography{urquhart_2021}

\end{document}